%% file: jmaster_apvast.tex
\documentclass[journal]{IEEEtran}
\ifCLASSINFOpdf
\else
\fi
\usepackage{amsmath}
\usepackage{array}
\usepackage{multirow}

\ifCLASSOPTIONcompsoc
 \usepackage[caption=false,font=normalsize,labelfont=sf,textfont=sf]{subfig}
\else
 \usepackage[caption=false,font=footnotesize]{subfig}
\fi
\usepackage{url}

\usepackage{cite}

\usepackage{amsmath,graphicx}
\usepackage{amssymb,amsfonts}
\usepackage{xcolor}

\usepackage{tikz}
\definecolor{mycolor1}{rgb}{0.00000,0.44700,0.74100}%

\usetikzlibrary{shapes.geometric, arrows, backgrounds}

\usepackage{pgfplots} 
\usetikzlibrary{plotmarks}
\pgfplotsset{compat=newest, plot coordinates/math parser=false, 
        every axis/.append style={
		label style={font=\small},
		tick label style={font=\small}
}}

\usepackage{pgfplotstable}
\usepackage{booktabs}

\pgfplotstableset{
  every head row/.style={before row=\toprule,after row=\midrule},
  every last row/.style={after row=\bottomrule},
}

\usepackage[caption=false,font=footnotesize]{subfig}

\DeclareMathOperator*{\argmin}{arg\,min}

\newcommand{\cc}[1]{{\color{black}#1}}

\newcommand{\vt}[1]{\boldsymbol{#1}}

\newcommand{\bt}[1]{{\color{blue}#1}}
\newcommand{\bbmtx}{\begin{bmatrix}}
\newcommand{\ebmtx}{\end{bmatrix}}

\newcommand{\mjr}[1]{{\color{black}#1}}
\newcommand{\rr}[1]{{\color{black}#1}}

\usepackage{booktabs,ragged2e}
\usepackage[flushleft]{threeparttable}

\newcommand\copyrighttext{%
  \footnotesize \textcopyright \ 2020 IEEE. Personal use of this material is permitted. Permission from IEEE must be obtained for all other uses, in any current or future media, including reprinting/republishing this material for advertising or promotional purposes, creating new collective works, for resale or redistribution to servers or lists, or reuse of any copyrighted component of this work in other works. DOI: 10.1109/TASLP.2020.3013397}
\newcommand\copyrightnotice{%
\begin{tikzpicture}[remember picture,overlay]
\setlength{\fboxrule}{0.0pt}
\node[anchor=south,yshift=5pt] at (current page.south) {\fbox{\parbox{\dimexpr\textwidth-\fboxsep-\fboxrule\relax}{\copyrighttext}}};
\end{tikzpicture}%
}

\begin{document}
\raggedbottom
%
\title{Signal-Adaptive and Perceptually Optimized Sound Zones with Variable Span Trade-Off Filters}
%
%
%

\author{Taewoong~Lee,~\IEEEmembership{Student Member,~IEEE,}
        Jesper~Kj\ae r~Nielsen,~\IEEEmembership{Member,~IEEE,}
        and~Mads~Gr\ae sb\o ll~Christensen,~\IEEEmembership{Senior Member,~IEEE}
\thanks{T. Lee, J. K. Nielsen, and M. G. Christensen are with the Audio Analysis Lab, CREATE, Aalborg University, 9000 Aalborg, Denmark (e-mail: \texttt{tlee, jkn, mgc@create.aau.dk})}}


%
%

\markboth{Journal of \LaTeX\ Class Files,~Vol.~14, No.~8, August~2015}%
{Shell \MakeLowercase{\textit{et al.}}: Bare Demo of IEEEtran.cls for IEEE Journals}
%



\maketitle

\copyrightnotice

\begin{abstract}
\input{"chapters/00_abstract/abstract.tex"}
\end{abstract}


%
\IEEEpeerreviewmaketitle

\section{Introduction}
%
%
%
%

\input{chapters/01_intro/ch1.tex}


\input{chapters/02_pvast/ch2.tex}

\input{chapters/03_apvast/ch3.tex}


\input{chapters/04_simulations/ch4_0_sims.tex}

\input{chapters/04_simulations/ch4_1_comp_complexity.tex}
\input{chapters/05_listeningtest/ch5_listening_test.tex}
\input{chapters/06_conclusion/ch6_conclusion.tex}

\appendices
\section{}
\label{appendix:ACproof}
\input{chapters/07_appendix/appendixA.tex}



\section*{Acknowledgment}
\mjr{The authors would like to thank Dr.-Ing. M.~Schneider and Assoc.~Prof. E.~A.~P.~Habets for providing the RIRs.}



\ifCLASSOPTIONcaptionsoff
  \newpage
\fi

\bibliographystyle{IEEEtran}
\bibliography{IEEEabrv_tlee,myabrv,references}

\end{document}

%% file: chapters/00_abstract/abstract.tex
Creating sound zones has been an active research field since the idea was first proposed. So far, most sound zone control methods rely on either an optimization of physical metrics such as acoustic contrast and signal distortion or a mode decomposition of the desired sound field. By using these types of methods, approximately $15$~dB of acoustic contrast between the reproduced sound field in the target zone and its leakage to other zone(s) has been reported in practical set-ups, but this is typically not high enough to satisfy the people inside the zones. In this paper, we propose a sound zone control method \rr{shaping} the leakage errors so that they are as inaudible as possible for a given acoustic contrast. The shaping of the leakage errors is performed by taking the time-varying input signal characteristics and the human auditory system into account when the loudspeaker control filters are calculated. We show how this \rr{shaping} can be performed using variable span trade-off filters, and we show theoretically how these filters can be used for trading signal distortion in the target zone for acoustic contrast. \rr{The proposed method is evaluated based on physical metrics such as acoustic contrast and perceptual metrics such as STOI. The computational complexity and processing time of the proposed method for different system set-ups are also investigated. Lastly, the results of a MUSHRA listening test are reported. The test results show that the proposed method provides more than $20\%$ perceptual improvement compared to existing sound zone control methods.}

%% file: chapters/01_intro/ch1.tex
\IEEEPARstart{S}{ound} zones are different listening areas in the same acoustic environment for different audio contents, and these zones are created by controlling a set of loudspeakers. Typically, two types of sound zones are considered: a bright zone and a dark zone. The bright zone is a confined region in which \rr{the} desired sound field is reproduced as faithfully as possible, whereas the dark zone is a confined region in which the energy of a reproduced sound field is suppressed as much as possible. These two zones are created by filtering the signals fed into the loudspeakers, and multiple bright zones can be obtained by superimposing the individual bright and dark zones for every input signal. Many different applications of sound zones have been studied\rr{,} including outdoor concerts \cite{brunskog2019full}, automobile cabins \cite{Cheer2013a, choi2015subband, Samarasinghe2016, choi2019, Widmark2019}, pedestrian alert systems \cite{quinn2014}, mobile devices \cite{Cheer2013},  personal computer\rr{s} \cite{Chang2009a}, and other applications \cite{jungmin2010ica, SGalvez2014}. 

Many different methods for designing the loudspeaker control filters have been proposed over the last two decades since the concept was first introduced in \cite{druyvesteyn1997personal}. Generally, these control methods seek to reproduce the desired sound field in the bright zone as faithfully as possible while also suppressing its leakage to the dark zone as much as possible. The proposed methods can be largely divided into three categories: mode matching methods, acoustic contrast control (ACC) methods, and pressure matching (PM) methods. Mode matching methods are based on that any sound field can be decomposed as an infinite sum of spatial harmonics. In practice, however, the sum is truncated up to a finite number of spatial harmonics often referred to as modes. The fundamental idea is based on \cite{Wu2011}, and several subsequent mode matching methods have been proposed \cite{Zhang2016jasa, Zhang2017, Jin2013, Donley2018}.

The ACC methods are designed to maximize the acoustic contrast, defined as the ratio of the acoustic potential energies between the bright and dark zones, and this is achieved by solving a generalized eigenvalue problem \cite{choi2002generation}. Since ACC only optimizes the acoustic contrast, it will in general not maintain the spatial characteristics of the desired sound field. Consequently, the ACC methods are most useful in situations where the spatial characteristics are either not important or very hard to reproduce due to complicated, dynamic acoustics environments such as in car cabins \cite{Cheer2013a, choi2015subband, Samarasinghe2016, choi2019, Widmark2019}. Various variations of the ACC method have been proposed. These include the energy difference maximization \cite{Shin2010a}, the planarity control \cite{Coleman2014planarity}, subband optimization \cite{choi2015subband}, multiple constraints on the acoustic contrast for different frequency bands \cite{Widmark2019}, and the broadband ACC (BACC) method \cite{Elliott2011us}. The BACC method is different from the other methods in that it operates in the time-domain instead of the frequency-domain. Since the BACC method typically will produce control filters that will filter out most of the energy in the input signal, except for the few frequencies where the maximum acoustic contrast can be obtained, the reproduced sound field will typically be severely distorted. Various ways of mitigating this problem have been proposed in \cite{Cai2013, Cai2014a, Schellekens2016}.

The PM methods produce control filters that minimize the reproduction error,  defined as the difference between the reproduced and desired sound fields in the bright and dark zones. The original method was proposed in \cite{Kirkeby1993, Poletti2008}. Compared to the ACC methods, the signal distortion is much smaller, but so is the acoustic contrast. To allow the user to trade-off these two, largely two types of combination method have been studied. In \cite{Moller2012}, a method which combines the energy difference maximization method \cite{Shin2010a} and PM \cite{Poletti2008} in order to control the acoustic contrast and the reproduction error has been proposed. In \cite{Chang2012}, a method sometimes referred to as ACC-PM has been proposed, and it is a more flexible PM method where the user can control the relative importance of reproducing the desired sound field and minimizing acoustic potential energy in the dark zone. The ACC-PM method has also been proposed in a broadband version in \cite{Simon-Galvez2015}. We note in passing that, despite its name, ACC-PM is actually not a combination of the ACC and PM methods and is also referred to by different names, e.g., in \cite{Moller2016, heuchel2018sound}. A true combination of the BACC and PM methods in the time-domain has recently been proposed in \cite{Lee2018}. 


Until now, an acoustic contrast of more than approximately $15$~dB has only been reported in highly idealized experiments where, e.g., an impractical number of loudspeakers are used, the acoustic environment is time-invariant, or the performance is evaluated using oracle knowledge of the acoustic environment \cite{Zhang2017}. Unfortunately, however, a much higher contrast than $15$~dB is needed, \mjr{as reported in \cite{Druyvesteyn1994}}. In \cite{Ramo2016, ramo2017validating}, it was \mjr{also} found that a target-to-interferer ratio (TIR) of at least $25$~dB is needed. TIR is a metric closely related to the acoustic contrast, but it measures the ratio of either the acoustic potential energy or loudness between the reproduced and interfering sound fields in a given zone (see \cite{Francombe2013a} for more on this).

Except for \cite{Donley2018} where the sound zones were optimized for preserving speech privacy and for \cite{Buerger2017a} where the pre-echoes were controlled over the attenuation of reflections in a reverberant environment, existing sound zone control methods design the control filters by minimizing physical metrics. A problem of quantifying the performance using physical metrics such as acoustic contrast and signal distortion is that they do not directly relate to the human auditory system. Moreover, the loudspeaker control filters are typically designed assuming input signals with flat spectra. The main advantage of this is that the control filters can be designed offline, but the disadvantage is that \mjr{array effort}\footnote{\mjr{The array effort is defined as the sum of mean squared control filters \cite{Elliott2012}.}} is wasted on controlling input signal frequency components which might not be present in the input signal or are inaudible. This is a general disadvantage of the frequency-domain methods in which the control filters are designed independently for every frequency bin. With the exception of \cite{Schneider2019}, sound zone control methods in the frequency-domain do not trade-off the reproduction error in one frequency bin for the reproduction error in another frequency bin.

In this paper, we propose \mjr{a perceptually optimized sound zone control method in the time-domain, which takes both the input signal characteristics and the human auditory system into account on a segment-by-segment level and gives explicit control of the trade-off between (weighted) acoustic contrast and signal distortion. This approach} is inspired by perceptual audio coding, where quantization errors have been successfully hidden by exploiting the characteristics of the human auditory system. Famously in the early 1990s \cite{brandenburg1992nmr}, the so-called $13$~dB miracle \cite[Ch.~10]{Bosi2003Kluwer} demonstrated that this approach drastically lowered the requirements to the signal-to-quantization noise level without impacting the perceived quality, and these principles have later been standardized in, e.g., MPEG-1/2 Layer-3 (MP3) \cite{brandenburg1994isompeg, Brandenburg2013}. In the sound zones application, we have reproduction errors instead of quantization errors. By using masking curves for designing weighting filters that shape the reproduction errors in a perceptually meaningful way, we can, therefore, ensure that the largest control effort is spent on maximizing the contrast and/or minimizing the reproduction error in the perceptually most important frequency regions. The proposed sound zone control method will be based on the variable span linear filter, which is a subspace approach initially proposed for signal enhancement \cite{Ephraim1995, Jensen1995, Doclo2002, vslf2016ieee, Benesty2016} \mjr{(see \cite{Jesper2018} for more on the relation between these problems). An interesting feature of the proposed method is that it reduces to existing sound zone control methods, such as broadband PM, BACC, and broadband ACC-PM, in special cases. We remark that this paper is an extension of our  preliminary work reported in \cite{Lee2018}, which considered only the un-weighted case, and\cite{Lee2019}, which considered the weighted but non-adaptive case. Moreover, we here also report more elaborate experimental analyses and results, including a MUSHRA (MUltiple Stimuli with Hidden Reference and Anchor) listening test \cite{itu15343mushra}.}

The paper is organized as follows: in Sec.~\ref{sec:wvastframework}, the sound zone control method with an arbitrary weighting of the reproduction error is explained, and it is shown how the input signal characteristics can be taken into account. In Sec.~\ref{sec:apvast}, we discuss how the weighting filters are designed to take the characteristics of the human auditory system into account. Furthermore, it is extended to explain how the input signals are segmented in blocks and how the loudspeaker control filters are updated. In Sec.~\ref{sec:simulations}, the performance of the proposed method is evaluated via not only typical physical metrics such as the acoustic contrast (AC), the normalized signal distortion (nSDP), and the TIR, but also perceptual metrics\rr{,} including the short-time objective intelligibility (STOI) \cite{Taal2011} and the instantaneous perceptual similarity measure (PSMt) from \mjr{the perception model based audio quality assessment method (PEMO-Q)}\footnote{\mjr{PEMO-Q was chosen because it shows a higher prediction accuracy in known data and a more robust prediction performance on completely new data over the perceptual evaluation of audio quality (PEAQ) in \cite{itu13871peaq}.}} \cite{Huber2006}. \mjr{In addition to this, the results of a MUSHRA listening test are reported.} Finally, in Sec.~\ref{sec:conclusion}, the paper is concluded.

%% file: chapters/02_pvast/ch2.tex
\section{A Weighted VAST Framework}
\label{sec:wvastframework}
\begin{figure}[t]
	\centering
	\includegraphics[width=8.6cm]{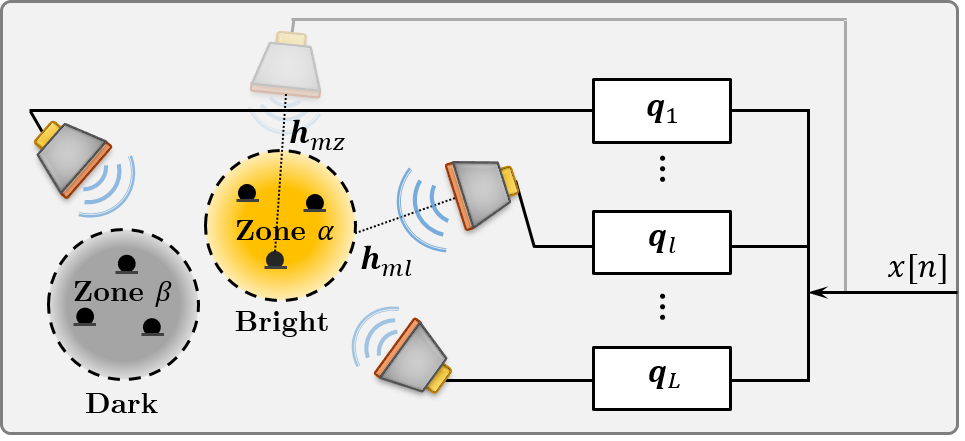}
	\caption{An illustration of a system geometry of sound zones. The input signal $x[n]$ is fed into $L$ loudspeakers after being filtered by the corresponding control filter $\{\vt{q}_l\}_{l=1}^{L}$. The RIR from loudspeaker $l$ to control point $m$ and the impulse response from virtual source $z$ to control point $m$ is represented as $\vt{h}_{ml}$ and $\vt{h}_{mz}$, respectively.} 
	\label{fig:soundzonesystem}
\end{figure}

In this section, the proposed weighted variable span trade-off (VAST) framework is described. To do this, we consider the simple system setup depicted in Fig.\ \ref{fig:soundzonesystem}. The figure shows the bright and dark zones as spatially confined regions sampled by $M_\text{B}$ and $M_\text{D}$ microphone positions, respectively. Moreover, the figure shows $L$ loudspeakers, with the $l$th loudspeaker having the finite impulse response (FIR) control filter with filter coefficients $\vt{q}_l$ and the input signal $x[n]$. As alluded to in the introduction, we can design multiple bright zones by superimposing the solutions to the individual bright and dark zones for each input signal. Throughout the theoretical part of this paper, we, therefore, consider the problem of creating a bright zone and a dark zone, and we use subscripts $_\text{B}$ and $_\text{D}$ to represent the bright and dark zones, respectively. The reproduced sound pressure $p_m[n]$ at microphone position or control point $m$ is represented by the linear convolution between input signal $x[n]$, the $L$ control filters $\{\vt{q}_l\}_{l=1}^{L}$ of length $J$, and the $L$ room impulse responses (RIRs) $\{\vt{h}_{ml}\}_{l=1}^L$ of length $K$, i.e.,
\begin{align}
	p_m[n] &= \sum_{l=1}^{L} \sum_{k=0}^{K-1} \sum_{j=0}^{J-1} x[n-k-j]h_{ml}[k]q_l[j]\notag\\
			&= \sum_{l=1}^{L}\vt{y}_{ml}^T[n]\vt{q}_l =  \vt{y}_{m}^T[n]\vt{q}, \label{eq:reprodpre}
\end{align}
where
\begin{align}
	\vt{y}_{ml}[n] &= \vt{X}[n]\vt{h}_{ml},\label{eq:yml_def}\\
	\vt{h}_{ml} &= \bbmtx h_{ml}[0] & \cdots & h_{ml}[K-1]\ebmtx^T,\\
	\vt{X}[n] &= \bbmtx x[n] & \cdots & x[n-K+1]\\
						\vdots & \ddots & \vdots\\
						x[n-J+1] & \cdots & x[n-K-J+2]\ebmtx,\\
	\vt{y}_m[n] &= \bbmtx \vt{y}_{m1}^T[n] & \cdots & \vt{y}_{mL}^T[n]\ebmtx^T,\\
	\vt{q} &= \bbmtx \vt{q}_1^T & \cdots & \vt{q}_L^T\ebmtx^T,\\
	\vt{q}_{l} &= \bbmtx q_{l}[0] & \cdots & q_{l}[J-1]\ebmtx^T.
\end{align}
The known signal vector $\vt{y}_{ml}[n]$ is the uncontrolled reproduced sound pressure at control point $m$ originating from loudspeaker $l$, as this is what we have when there is no control over the zones, i.e., the control filters are all equal to the Kronecker delta function. The goal is then to design the control filters $\vt{q}$ so that the reproduced sound pressure $p_m[n]$ matches a desired sound pressure $d_m[n]$ across all control points as well as possible. Typically, the desired pressures are all $0$ for the control points in the dark zone, whereas those in the bright zone are defined as part of a sound field generated by a virtual source $z$ emitting $x[n]$. Thus, the desired sound pressure at control point $m$ is defined as
\begin{align}
	d_m[n] &= 
		\begin{cases}
			(h_{mz} * x)[n] & m\in \mathcal{M}_\text{B}\\
			0 & m\in \mathcal{M}_\text{D}
		\end{cases},
		\label{eq:despre}
\end{align}
where $*$ denotes the linear convolution operator, $\mathcal{M}_\text{B}$ and $\mathcal{M}_\text{D}$ are the set of control point indices for the bright and dark zones, respectively, and $h_{mz}[n]$ is the impulse response from the virtual source $z$ to control point $m$\rr{,} as depicted in Fig.~\ref{fig:soundzonesystem}. Note that sound zone control methods have to implicitly perform dereverberation in order to match the desired and reproduced sound fields if the desired sound field is defined in an anechoic environment.

In sound zone control, two zones labeled $\alpha$ and $\beta$ are typically considered as illustrated in Fig.\ \ref{fig:soundzonesystem}. If we consider two zones, each having their own desired sound field, then the bright and dark zones for audio input signal $x^{(\alpha)}[n]$ are zone $\alpha$ and zone $\beta$, respectively, and those for audio input signal $x^{(\beta)}[n]$ are zone $\beta$ and zone $\alpha$, respectively. To this end, multiple bright zones can be obtained when the two reproduced sound fields are superposed. 

How close the reproduced sound field is to the desired sound field can be quantified by the reproduction error, defined as the difference between the desired and reproduced sound pressures across all control points in a given zone such that
\begin{align}
	\varepsilon_m[n] &= d_m[n] - p_m[n]\ .
	\label{eq:re}
\end{align}
More generally, we can filter the reproduction error by a weighting filter $w_m[n]$ so that
\begin{align}
	\tilde{\varepsilon}_m[n] &= (w_m*\varepsilon_m)[n] = \tilde{d}_m[n]-\tilde{p}_m[n]\ ,
	\label{eq:wre}
\end{align}
where, e.g., $\tilde{p}_m[n]$ means that $p_m[n]$ has been filtered with the weighting filter $w_m[n]$. If we plug this into \eqref{eq:reprodpre}, we obtain the weighted and reproduced sound pressure at control point $m$ as
\begin{align}
	\tilde{p}_m[n] = (w_m*p_m)[n] = \sum_{l=1}^{L}\tilde{\vt{y}}_{ml}^T[n]\vt{q}_l =  \tilde{\vt{y}}_{m}^T[n]\vt{q}
\end{align}
where $\tilde{\vt{y}}_{ml}^T[n]$ is defined as in \eqref{eq:yml_def}, except for that the source signal $x[n]$ is pre-filtered with the weighting filter. Note that the weighting filter is assumed to be known and is used to shape the reproduction error according to some design criterion. This will be elaborated upon in the next section.

We are now able to measure the distance between $\tilde{p}_m[n]$ and $\tilde{d}_m[n]$ for each of the zones. This can describe how much distortion is present in the bright zone and how much power is remaining in the dark zone. This allows us to define the weighted signal distortion power (SDP)  $\tilde{\mathcal{S}}_B(\vt{q})$ and the weighted residual error power $\tilde{\mathcal{S}}_D(\vt{q})$, respectively, as
\begin{align}
	\tilde{\mathcal{S}}_\text{B}(\vt{q}) &= \frac{1}{|\mathcal{M}_\text{B}|N}\sum_{n=0}^{N-1}\sum_{m\in\mathcal{M}_\text{B}} |\tilde{\varepsilon}_m[n]|^2\notag\\
		&= \tilde{\sigma}_d^2-2\vt{q}^T\tilde{\vt{r}}_\text{B}+\vt{q}^T\tilde{\vt{R}}_\text{B}\vt{q}, \label{eq:errbr}\\
	\tilde{\mathcal{S}}_\text{D}(\vt{q}) &= \frac{1}{|\mathcal{M}_\text{D}|N}\sum_{n=0}^{N-1}\sum_{m\in\mathcal{M}_\text{D}} |\tilde{\varepsilon}_m[n]|^2 = \vt{q}^T\tilde{\vt{R}}_\text{D}\vt{q},\label{eq:errdk}
\end{align}
where $N$ is the number of observations and
\begin{align}
	\tilde{\sigma}_d^2 &= \frac{1}{|\mathcal{M}_\text{B}|N}\sum_{n=0}^{N-1}\sum_{m\in\mathcal{M}_\text{B}} |\tilde{d}_m[n]|^2, \label{eq:sd}\\
	\tilde{\vt{r}}_\text{B} &= \frac{1}{|\mathcal{M}_\text{B}|N}\sum_{n=0}^{N-1}\sum_{m\in\mathcal{M}_\text{B}} \tilde{\vt{y}}_m[n]\tilde{d}_m[n], \label{eq:rB}\\ 
	\tilde{\vt{R}}_\text{C} &= \frac{1}{|\mathcal{M}_\text{C}|N}\sum_{n=0}^{N-1}\sum_{m\in\mathcal{M}_\text{C}} \tilde{\vt{y}}_m[n]\tilde{\vt{y}}_m^T[n],\label{eq:RBRD}
\end{align}
with $\tilde{\vt{R}}_\text{C}$ for $\text{C} \in \{\text{B}, \text{D}\}$ being the spatial correlation matrix for the corresponding zone, $\tilde{\vt{r}}_\text{B}$ being the spatial correlation vector for the bright zone, and $\tilde{\sigma}_d^2$ being the variance of the desired sound field.

Using these definitions, we can pose the convex optimization problem
\begin{align}
    \text{minimize } \tilde{\mathcal{S}}_\text{B}(\vt{q}) \text{ subject to } \tilde{\mathcal{S}}_\text{D}(\vt{q}) \leq \epsilon, \label{eq:cvxopt}
\end{align}
where $\epsilon$ is a nonnegative scalar representing the power allowed in the dark zone. The Lagrangian function corresponding to the problem in \eqref{eq:cvxopt} is
\begin{align}
	\mathcal{L}(\vt{q}) &= \tilde{\mathcal{S}}_\text{B}(\vt{q}) + \mu (\tilde{\mathcal{S}}_\text{D}(\vt{q})-\epsilon),
	\label{eq:costfcn}
\end{align}
where $\mu \geq 0$ is the Lagrange multiplier. As also assumed in the signal enhancement literature \cite{vslf2016ieee}, this Lagrange multiplier is here treated as a user-defined parameter that controls the trade-off between minimizing $\tilde{\mathcal{S}}_\text{B}(\vt{q})$ and suppressing $\tilde{\mathcal{S}}_\text{D}(\vt{q})$. We note in passing that minimizing \eqref{eq:costfcn} for $\mu=1$ and $\mu$ equal to a constant produces the broadband PM solution and the broadband ACC-PM solution, respectively, when no weighting is applied, i.e., $w[n]$ is the Kronecker delta function, and the input signal is assumed to have a flat spectrum.

The matrices $\tilde{\vt{R}}_\text{B}$ and $\tilde{\vt{R}}_\text{D}$ are real, symmetric, and at least semi-positive definite matrices. Provided that $\tilde{\vt{R}}_\text{D}$ is positive definite, these properties allow us to compute a joint diagonalization for those two matrices \cite{Jesper2018}, \cite[Ch.~8.7]{Golub2013matrix}. As exploited for signal enhancement \cite{vslf2016ieee}, we can use this diagonalization to obtain more control over the trade-off between the SDP and the acoustic contrast. Specifically, we obtain this control by solving \eqref{eq:costfcn} with a low-rank approximation to the control filters $\vt{q}$. The two matrices $\tilde{\vt{R}}_\text{B}$ and $\tilde{\vt{R}}_\text{D}$ can be jointly diagonalized as
\begin{align}
	\vt{U}_{LJ}^T\tilde{\vt{R}}_\text{B}\vt{U}_{LJ} &= \vt{\Lambda}_{LJ}, & \vt{U}_{LJ}^T\tilde{\vt{R}}_\text{D}\vt{U}_{LJ} &= \vt{I}_{LJ}
	\label{geigproblem},
\end{align}
where $\vt{\Lambda}_{LJ} = \mathrm{diag}(\lambda_1,\cdots,\lambda_{LJ})$ is a diagonal matrix containing the generalized eigenvalues in descending order, i.e., $\lambda_1 \geq \cdots \geq \lambda_{LJ} \geq 0$, $\vt{I}_{LJ}$ is the $LJ\times LJ$ identity matrix, and $\vt{U}_{LJ}$ is a nonsingular matrix containing the generalized eigenvectors sorted according to the eigenvalues. The matrices $\vt{U}_{LJ}$ and $\vt{\Lambda}_{LJ}$ are computed by solving the eigenvalue problem
\begin{align}
    \tilde{\vt{R}}_\text{D}^{-1}\tilde{\vt{R}}_\text{B}\vt{U}_{LJ} &= \vt{U}_{LJ}\vt{\Lambda}_{LJ}\ .
\end{align}
\noindent It is worth noting that $\tilde{\vt{R}}_\text{D}$ is typically positive definite when $M_D\text{min}(N, K+J-1) \geq LJ$.

Since any vector can be represented as a linear combination of the columns of a nonsingular matrix, $\vt{q}$ can be written as
\begin{align}
    \vt{q} &= \vt{U}_{LJ}\vt{a}_{LJ},\label{eq:q_lincomb_LJ}
\end{align}
where $\vt{a}_{LJ}$ is an $LJ$ coefficient vector. If we plug \eqref{eq:q_lincomb_LJ} into \eqref{eq:errbr} and \eqref{eq:errdk}, we obtain 
	\begin{align}
	\tilde{\mathcal{S}}_\text{B}(\vt{U}_{LJ}\vt{a}_{LJ}) &= \tilde{\sigma}_d^2-2\vt{a}_{LJ}^T\vt{U}_{LJ}^T\tilde{\vt{r}}_\text{B} + \vt{a}_{LJ}^T\vt{\Lambda}_{LJ}\vt{a}_{LJ},\label{eq:errbr_Ua}\\
	\tilde{\mathcal{S}}_\text{D}(\vt{U}_{LJ}\vt{a}_{LJ}) &= \vt{a}_{LJ}^T\vt{a}_{LJ}.\label{eq:errdk_Ua} 
	\end{align}
Interestingly, we can observe from \eqref{eq:errdk_Ua} that $\tilde{\mathcal{S}}_\text{D}$ is only represented by $\vt{a}_{LJ}$. Hence, this joint diagonalization leads us to analyze how $\tilde{\mathcal{S}}_\text{B}$ and $\tilde{\mathcal{S}}_\text{D}$ behave in terms of the eigen information. Furthermore, we benefit from introducing a $V(\leq LJ)$-rank approximation by forcing the $LJ-V$ smallest eigenvalues to $0$, which directly reduces $\tilde{\mathcal{S}}_\text{D}$. How this affects $\tilde{\mathcal{S}}_\text{B}$ is explained later. Now we can approximate $\vt{q}$ by using the first $V$ eigenvectors such that
\begin{align}
	\vt{q} &\approx \vt{U}_{V}\vt{a}_V,\label{eq:q_lincomb_v}
\end{align}
where $1\leq V \leq LJ$ and optimize $\mathcal{L}$ over $\vt{a}_V$ instead of $\vt{q}$ directly. The cost function \eqref{eq:costfcn} is then
\begin{equation}
    \begin{aligned}
        \mathcal{L}(\vt{U}_{V}\vt{a}_V) &= \tilde{\sigma}_d^2-2\vt{a}_{V}^T\vt{U}_{V}^T\tilde{\vt{r}}_\text{B} + \vt{a}_{V}^T\vt{\Lambda}_{V}\vt{a}_{V} \\ &\phantom{abc} + \mu (\vt{a}_{V}^T\vt{a}_{V}-\epsilon).
        \label{eq:costfcn_v}
    \end{aligned}
\end{equation}
The solution to this is analytically derived and given by
\begin{align}
	\vt{a}_\textup{P-VAST}(V,\mu) &= \argmin_{\vt{a}_\textup{V}} \mathcal{L}(\vt{U}_{V}\vt{a}_\textup{V}) \nonumber\\ 
		&= \left[\vt{\Lambda}_V+\mu\vt{I}_V\right]^{-1}\vt{U}_V^T\tilde{\vt{r}}_\text{B}\ .	\label{eq:pvast_coefficient}
\end{align}
Finally, we plug \eqref{eq:pvast_coefficient} into \eqref{eq:q_lincomb_v} and obtain the control filter as
\begin{align}
	\vt{q}_\textup{P-VAST}(V,\mu) = \vt{U}_{V}\vt{a}_\textup{P-VAST}(V,\mu) = \sum_{v=1}^V \frac{\vt{u}_v^T\tilde{\vt{r}}_\text{B}}{\lambda_v+\mu}\vt{u}_v,
	\label{eq:pvast_confilter}
\end{align}
where $\lambda_v$ and $\vt{u}_v$ are the $v$th generalized eigenvalue and eigenvector, respectively.

\input{figures/fig2_vmu_plane.tikz}

\begin{figure*}[t!]
	\centering
	\includegraphics[width=17.2cm]{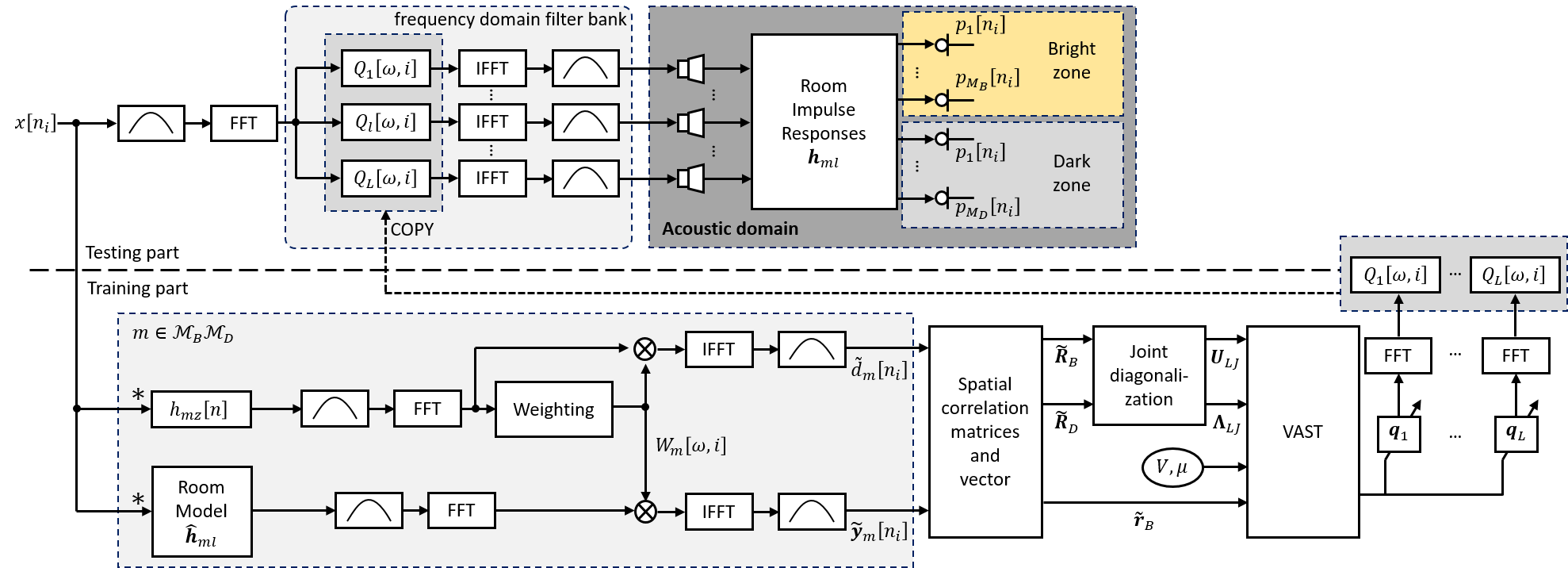}
	\caption{The block diagram to get the reproduced sound field at the given time segment $i$ \mjr{for AP-VAST}. Note that * denotes the linear convolution in the time-domain, $W_m[\omega,i]$ denotes the weighting filter in the frequency-domain at frequency $\omega$ and time segment $i$ at control point $m$, FFT and IFFT denote the fast Fourier transform and its inverse, respectively, $\mathcal{M}_\text{B}\mathcal{M}_\text{D}$ denotes a concatenated index set of $\mathcal{M}_\text{B}$ and $\mathcal{M}_\text{D}$, and $Q_l[\omega,i]$ denotes the control filters in the frequency-domain at frequency $\omega$ and time segment $i$ for loudspeaker $l$.}
	\label{fig:blockdiagram_pvast}
\end{figure*}

Interestingly, we can obtain different solutions by varying $V$ and $\mu$, including many existing solutions as shown in Fig. \ref{fig:plane_vmu_vast} assuming no weighting of the reproduction error and an input signal with a flat spectrum. For example, the BACC solution is obtained when $V=1$, the broadband PM (or Wiener) solution is obtained when $V=LJ$ and $\mu=1$, the broadband ACC-PM solution is obtained when $V=LJ$, and the MVDR solution is obtained when $V=LJ$ and $\mu=0$. As we have shown in App.~\ref{appendix:ACproof}, the maximum acoustic contrast but also the maximum SDP is obtained for $V=1$. Increasing $V$ and keeping $\mu$ fixed decrease both, and we obtain the minimum SDP but also the minimum acoustic contrast for the maximum number of eigenvectors, i.e., $V=LJ$. Thus, $V$ is also a user parameter that can be used for controlling the trade-off between the acoustic contrast and the SDP. Clearly, $\mu$ also controls aspects of this trade-off and can, e.g., be set so that the acoustic contrast is completely ignored (the MVDR solution). 



%% file: figures/fig2_vmu_plane.tikz
\begin{figure}[t!]
	\centering
	\begin{tikzpicture}[scale=1]
	\pgfplotsset{footnotesize, compat=newest}
	\draw[solid,-] (0,0) -- (0,2.2) node [anchor=	south east] {$\mu$};
	
	\filldraw [fill=mycolor1!30!white, draw=mycolor1!30!white] (0,0) rectangle (7.2,2.2) node[pos=0.5] {\Large{\textbf{VAST}}};
	
	\draw (0, 1pt) -- (0,-1pt) node[anchor=north] {$1$};
	\draw (7.2, 1pt) -- (7.2,-1pt) node[anchor=north] {$LJ$};
	\draw (1pt, 0) -- (-1pt,0) node[anchor=east] {$0$};
	\draw (1pt, 1.5) -- (-1pt,1.5) node[anchor=east] {$1$};
	
	\draw[line width=1.2pt, solid,->] (0, 0) -- node[anchor=west] {\small BACC} ++ (0, 2.2);
	
	\draw[line width=1.2pt, solid,->] (7.2, 0) -- (7.2, 2.2);
	\draw[line width=1.2pt, solid,-] (7.2, 1) -- node[anchor=east] {\small BACC-PM} ++ (0, 0.1);
	
	\fill[black] (7.2, 1.5) circle (2pt);
	\draw[solid, ->] (7, 1.7) -- (7.1,1.6) node[anchor=south east] {\small PM, Wiener};
	
	\fill[black] (7.2, 0) circle (2pt);
	\draw[solid, ->] (7, 0.3) -- (7.1,0.1) node[anchor=south east] {\small MVDR};
	
	\draw[line width = 1.2pt, <->] (1.2, 1.5) -- node[above] {\small VS Wiener} ++ (5, 0);
	\draw[line width = 0.1pt, solid] (0, 1.5) -- (7.2, 1.5);
	
	\draw[line width = 1.2pt, <->] (1.2, 0) -- node[above] {\small VS MD} ++ (5, 0);
	\node[text width = 5cm] at (4.05,-0.3) {The number of eigenvectors $V$};
 	\draw[line width = 0.1pt, solid] (0, 0) -- (7.2, 0);
	\end{tikzpicture}
	\caption{VAST plane \rr{that} illustrates how the various special cases of the VAST solutions are related in terms of a function of the user parameters $1 \leq V \leq LJ$ and $\mu \geq 0$.}
	\label{fig:plane_vmu_vast}
\end{figure}

%% file: chapters/03_apvast/ch3.tex
\section{Signal-Adaptive and Perceptually Optimized Sound Zones}
\label{sec:apvast}

As alluded to in the introduction, audio coding was revolutionized by exploiting simple mathematical models for the human auditory system. These models encode the principle that a certain sound also known as maskee becomes less audible or inaudible in the presence of a stronger masker close to the maskee in the time- and/or frequency-domain \cite{Moore2013}. This phenomenon is generally referred to as the masking effect, and it allows us to make large modifications to audio signals without changing how they are perceived by humans. In the sound zones application, we can only find a set of control filters that renders the reproduction errors to exactly zero provided that the multiple-input/multiple-output inverse theorem (MINT) conditions are satisfied \cite{Miyoshi1988, Nelson1995}, i.e., it is necessary (but not sufficient) to have more loudspeakers than control points, something which is seldom satisfied in practice. We thus cannot avoid making a reproduction error, but we can seek to shape this error to be as inaudible as possible. In the proposed sound zone control method, the reproduction error is shaped in the following way. For control point $m$, we compute a masking curve from a given input signal based on a psychoacoustic model, e.g.,  \cite{VanDePar2005}. \mjr{This masking curve is defined as an amplitude spectrum describing the sound pressure level below which any sound is modeled to be inaudible. The weighting filter is calculated as the reciprocal of the masking curve.} In other words, we apply a small weight to those part of the spectrum where the masker has a high power, whereas we penalize reproduction errors more by applying a larger weight in those part of the spectrum where the masker has a low or no power.

\begin{table}[!t]
	\caption{Desired signal and masker for control point $m$ and given time segment $i$}
	\begin{center}
		{\renewcommand{\arraystretch}{1.7}
			\begin{tabular}{|c|c|c|}
				\hline
				\textbf{Zone} & $\alpha \quad (m\in\mathcal{M}_\alpha)$ & $\beta \quad (m\in\mathcal{M}_\beta)$\\
				\hline
				\textbf{Desired signal} & $d^{(\alpha)}_m[n_i]$ & $d^{(\beta)}_m[n_i]$\\
				\hline
				\textbf{Masker} & $d^{(\alpha)}_m[n_i]$ & $d^{(\beta)}_m[n_i]$\\
				\hline
		\end{tabular}}
		\label{tzonesetup}
	\end{center}
\end{table}

To compute the masking curve \mjr{at} control point $m$, we must first figure out where the control point is located. If it is in zone $\alpha$, say, the masking curve is computed from the desired signal $d_m^{(\alpha)}[n]$ at this control point when zone $\alpha$ is the bright zone. Note that we have used the superscript $(\cdot)^{(\alpha)}$ on the desired signal to stress that this signal does not change with which zones are considered as bright or dark zones. Thus, when zone $\alpha$ is considered to be the bright zone, the masking curves for the control points in the dark zone, i.e., zone $\beta$, are calculated not from $d_m[n]= 0$ but from $d_m^{(\beta)}[n]$. \mjr{Note that the masking curves are calculated from the desired signal to avoid an iterative procedure for computing the control filters, although the actual masker is the reproduced signal. For precisely this reason, the masking curves used in audio coding are also calculated from the unquantized signal, although the actual masker is the quantized signal \cite{Painter2000}.} The above discussion is summarized in Table~\ref{tzonesetup}. If a zone is desired to be a dark zone, the masking curve for the zone will simply be the threshold in quiet. 


Although average masking curves can be computed from audio signals, we can expect to obtain the best performance if the masking curves are updated on a segment basis so that the control filters are adapted to the current input signal segment. To do this, we divide $x[n]$ into $I$ time segments, and $\vt{q}$ is calculated at each of these time segments. For the segment-wise approach, the observation index $n$ can be considered as a local time-index, and this is related to the global time-index $n_i$ as
\begin{align}
	n_i &= (N-\eta)(i-1) + n, & i \in \mathcal{I}, \label{eq:obidx_i}
\end{align}
where $\mathcal{I}$ denotes the set of the segment indices, $\eta\in\{0,1,\ldots,N-1\}$ is the number of overlapping samples between segments, and $n = 0,1,\cdots,N-1$. \rr{This indexing is used in Fig.~\ref{fig:blockdiagram_pvast}, which shows the implementation of the proposed sound zone control method, referred to as signal-adaptive and perceptually optimized variable span trade-off (AP-VAST). Fig.~\ref{fig:blockdiagram_pvast} shows that the weighting and the filtering with the control filters are implemented in the short-time Fourier-transform (STFT) domain with a $50\%$ overlap} and with identical analysis and synthesis windows given by \cite{Malvar1990}
\begin{align}
    g[n] &= \sin{\bigg\{\frac{\pi}{N}\bigg(n+\frac{1}{2}\bigg)\bigg\}}\ . \label{eq:sinewindow}
\end{align}
This implementation of the time-varying filtering is adopted since it has proven successful in many speech and audio processing applications, including audio coding. The room model $\hat{\vt{h}}_{ml}$ indicates that the RIRs have been measured or modeled in advance. \mjr{It should be noted that AP-VAST has a special case, which we refer to as perceptually optimized sound zones (P-VAST), that uses averaged input signal statistics and masking curves.}

%% file: chapters/04_simulations/ch4_0_sims.tex
\section{Experimental Validation and Discussion}
\label{sec:simulations}
This section presents an evaluation of the proposed method and comparisons to the reference methods in both anechoic and reverberant environments. We here consider the case of two bright zones. In other words, two input signals are considered, and each of them becomes the desired signal in the corresponding zone.  \mjr{Individual problems of sound zones for each input signal are solved, then the reproduced sound fields are superimposed.}

\subsection{Set-up}
\label{ssec:sim_systemsetup}
\begin{figure}[t]
	\centering
	\includegraphics[width=6cm]{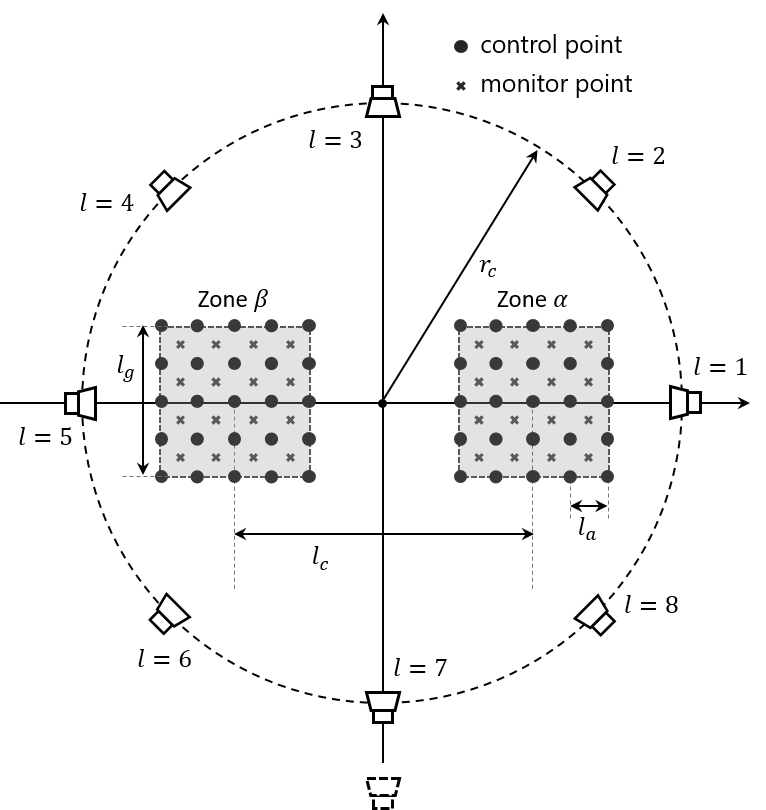}
	\caption{The system geometry used in the validations. Note that the illustration does not follow the actual scale but is magnified for better visualization.} 
	\label{fig:simsetup}
\end{figure}

As depicted in Fig.~\ref{fig:simsetup}, a circular array with a radius of $r_c = 2$~m with eight omnidirectional loudspeakers evenly placed on the circumference was considered. \rr{The two virtual sources were located outside of the array at the same location, which was $0.5$~m away from the $7$th loudspeaker. It is depicted as a dashed line loudspeaker in Fig. \ref{fig:simsetup}.} The zones were located in the interior of the loudspeaker array and spatially sampled by $25$ control points on a $2$D grid with $l_a = 5$~cm spacing between the control points to cover the size of a human head. The control points are shown as black dots in Fig.~\ref{fig:simsetup} and were used to calculate the control filters. Besides, $16$ monitor points were used to evaluate the performance. These points are shown as gray crosses in Fig.~\ref{fig:simsetup} and located in between the control points. The centers of the two zones were $l_c = 2$~m apart from each other. \rr{All loudspeakers, control points, and the virtual sources were assumed to be located in the same plane at the height of $1.5$~m.} All the parameters that are common in all experiments are summarized in Table~\ref{table:spdetails}. 

For the performance evaluation, the typical physical metrics -- AC, nSDP, and TIR -- as well as the perceptual metrics -- STOI \cite{Taal2011} and PSMt\footnote{PSM and PSMt showed a similar trend throughout all experiments, so only the results of PSMt are displayed.} \cite{Huber2006} -- were used. AC, nSDP, and TIR are here defined as 
\begin{align}
    \text{AC} &= 10 \log_{10}{\left( \frac{M_D}{M_B} \frac{ \displaystyle\sum_{n=0}^{N-1}{\displaystyle\sum_{m \in \mathcal{M}_B}{|p_m[n]|^2}}} { \displaystyle\sum_{n=0}^{N-1}{\displaystyle\sum_{m \in \mathcal{M}_D}{|p_m[n]|^2}}} \right)}, \label{eqref:def_ac}\\
    \text{nSDP}_m &= 10 \log_{10}{\bigg( \frac{\sum_{n=0}^{N-1}{|p_m[n] - d_m[n]|^2}}{\sum_{n=0}^{N-1}{|d_m[n]|^2}} \bigg)}, \label{eqref:def_sdp}\\
    \text{TIR}_m^{(\alpha; \beta)} &= 10 \log_{10}{\Bigg( \frac { \sum_{n=0}^{N-1}{|p_m^{(\alpha)}[n]|^2} } { \sum_{n=0}^{N-1}{|p_m^{(\beta)}[n]|^2} } \Bigg)}\ . \label{eqref:def_tir}
\end{align}
\rr{Although these metrics are typically defined for an input signal with a flat spectrum, we calculated the metrics from the desired and reproduced sound fields because they should depend on those sound fields.} \mjr{Since AC is defined as a zone-wise metric, AC was calculated zone-wise. On the other hand, nSDP, TIR, and the perceptual metrics were calculated point-wise. Note that the metrics calculated point-wise can be easily converted to the zone-wise metrics and vice versa. The mean values and the error bars with the $95\%$ confidence intervals are shown for the metrics calculated point-wise}\footnote{For the case of multiple bright zones, a family of perceptual source separation metrics in \cite{Emiya2011} was also reviewed. Unlike STOI and PSMt, however, these metrics did not correlate very well with an informal listening test, so we did not include the metrics in this paper.}. 

\begin{table}[!t]
	\caption{The Parameter Details for the Simulations}
	\begin{center}
		{\renewcommand{\arraystretch}{1.1}
			\begin{tabular}{ c | c | c | c }
			\toprule[\heavyrulewidth]\toprule[\heavyrulewidth]
				\textbf{Variable} & \textbf{Value} & \textbf{Variable} & \textbf{Value}\\
				\hline
				\hline
				$M_B$, $M_D$ (control) & $25$ & $M^\prime_B$, $M^\prime_D$ (monitor) & $16$\\
				\hline
				$L$ & $8$ & $r_c$ & $2$~m\\
				\hline
				$l_c$ & $2$~m & $l_g$ & $0.2$~m\\
				\hline
				$l_a$ & $0.05$~m\\
				\hline
				\hline
				sampling frequency & $16$~kHz & speed of sound, $c$ & $343$~m/s\\
				\hline
				$K$ & $3200$ & $J$ & $240$\\
				\bottomrule[\heavyrulewidth]
		\end{tabular}}
		\label{table:spdetails}
	\end{center}
\end{table}

It is also important to note that AC and nSDP are \rr{the} metrics per input signal, whereas TIR is the metric per zone. Besides, the reference signal for STOI and PSMt is the desired signal at each point $d_m^{(\alpha)}[n]$, and the processed signal for STOI and PSMt is the observed signal that is the sum of the reproduced signal $p_m^{(\alpha)}[n]$ and $p_m^{(\beta)}[n]$ at the corresponding point. We used the freely available \textsc{Matlab} toolboxes of STOI and PSMt for obtaining the results \cite{Taal2011, Huber2006}, respectively, and the default values in each toolbox were used. Finally, the RIRs were calculated using the RIR generator toolbox \cite{Habets2010room}, which is a \textsc{Matlab} implementation of the image source method \cite{Allen1979}, for both the anechoic environment ($T_{60} = 0$~s) and the reverberant environment ($T_{60} = 200$~ms). Lastly, the $48$ kHz RIRs were generated and then downsampled to $16$~kHz. 

As the reference methods for the performance comparison to the proposed method, AP-VAST, we used broadband PM (i.e., equivalent to broadband ACC-PM in \cite{Simon-Galvez2015} with $\xi = \mu/(1+\mu) = 0.5$), the frequency-domain ACC\footnote{Note that a regularization based on the truncated singular value decomposition (TSVD) in \cite{Chang2012} was used and that the magnitude normalization factor of the control filter was calculated as described in \cite{Chang2012} for the entire frequency range, except at the first frequency component (i.e., the DC-component) and the Nyquist frequency\rr{,} which were both set to $0$.} \cite{choi2002generation}, VAST \cite{Lee2018}, and P-VAST. As a baseline, we also evaluated the performance without any control, i.e., the control filters were all set equal to the Kronecker delta function. For AP-VAST, the time-varying weighting filters were obtained from masking curves computed from each $60$~ms time segment with \rr{a} $50\%$ overlap, whereas the weighting filter for P-VAST was calculated from an averaged masking curve computed from all $60 \text{ ms}$ time segments. The psychoacoustic model in \cite{VanDePar2005} was used to calculate masking curves. AP-VAST followed the implementation described in Fig.~\ref{fig:blockdiagram_pvast}. Lastly, a control filter length of $J = 240$ samples (i.e., $15$~ms at a sampling frequency of  $16$~kHz which corresponds to a frequency grid of $66.67$~Hz in the ACC method) for all methods\rr{,} and a segment length of $N = 960$ samples (i.e., $60$~ms at a sampling frequency of $16$~kHz) for AP-VAST were used\footnote{To compute the joint diagonalization, $\tilde{\vt{R}}_{D}$ has to be full rank, which requires $M_D\text{min}(N, K+J-1) \geq LJ$.}. Therefore, the number of eigenvectors $V$ was in the range from $1$ to $1920$ since $1\leq V \leq LJ$. Note that the same $(V, \mu)$ for the two input signals were used for each data point in the following figures. 

\input{chapters/04_simulations/exp0/exp0_summary_ztpia_msk_info.tex}

\subsection{Performance evaluation on the proposed method AP-VAST}
\label{ssec:sim1_performEvaluation}
In the first experiment, we considered two input signals as six seconds of dialogues excerpt from the Disney movie ``Zootopia'' in two different languages (English and Danish). These were used for the desired signal in zone $\alpha$ and zone $\beta$, respectively. The energy of the two signals was calibrated to be identical, and they were downsampled from $44.1$~kHz to $16$~kHz. The number of segments for AP-VAST was equal to $I = 201$. 

Masking curves of the speech signal for zone $\alpha$ are shown in Fig.~\ref{fig:exp0_msk_info}. Note that the desired signal at the $1$st control point $d_1^{(\alpha)}[n]$ is considered. As an example of the masking curves, one of the silent segments and one of the speech segments are depicted in blue and in red, respectively in panel~\protect\subref{subfig:exp0_signal}. From these segments, the masking curves corresponding to each of the segments are calculated and plotted in Fig.~\ref{fig:exp0_msk_info}~\protect\subref{subfig:exp0_sig_sil} with the same color as the segments. In other words, if the input segment barely contains any signal characteristics, a masking curve (blue) close to the threshold-in-quiet with a shallow slope is obtained. Otherwise, the masking curve (red) is calculated based on the corresponding input segment. Lastly, the masking curve (gray) from each segment and the averaged masking curve (black) across all the segments are shown in Fig.~\ref{fig:exp0_msk_info}~\protect\subref{subfig:exp0_allseg}. The averaged masking curve is used in P-VAST. 

\input{chapters/04_simulations/exp1/exp1_summary.tex}

First, AC and nSDP obtained by AP-VAST from the speech signal for zone $\alpha$ for five different $\mu$'s are shown in Fig.~\ref{fig:exp1_results}. Regardless of $\mu$, a clear trend is that AC and nSDP decrease with increasing $V$. For $V = 1$, in any case of $\mu$, the highest AC but also the largest nSDP is obtained. The smallest nSDP along with the lowest AC is obtained when $V = LJ$ with a fixed $\mu$. This \rr{trend} certainly shows the trade-off between AC and SDP\rr{,} which is the core property of AP-VAST. 

\input{chapters/04_simulations/exp2/exp2_summary.tex}

As $\mu$ increases from $\mu = 0$, both AC and nSDP increase for a fixed $V$. The lower bound of AC and nSDP can be observed when $\mu = 0$. In this case, AP-VAST searches for the solution to minimize the signal distortion in the bright zone instead of reducing the residual power in the dark zone. Hence, the case of $V = 1920$ and $\mu = 0$ guarantees the least nSDP, but so is AC, which is close to $0$ dB. Interestingly, AC and nSDP become less sensitive for $V$ as $\mu$ increases. If $\mu$ is large enough, e.g., $\mu = 100$, the degradation on AC and nSDP with increasing $V$ is smaller than when $\mu$ is small, e.g., $\mu = 1$. This \rr{trend} can be interpreted as AP-VAST seeks the solution to reduce the residual power in the dark zone more than to minimize the signal distortion in the bright zone as $\mu$ increases. Even though TIR is not plotted in this experiment, TIR follows the same trend as AC. 

\input{chapters/04_simulations/exp3/exp3_summary_phy.tex}

Secondly, STOI and PSMt obtained by AP-VAST from the speech signal for zone $\alpha$ for five different $\mu$'s are shown in Fig.~\ref{fig:exp2_results}. As alluded to earlier in this section, the observed signal is the sum of the reproduced signal by input signal $\alpha$ and the interference signal by input signal $\beta$. This affects STOI and PSMt to decrease\rr{,} especially as $\mu$ decreases and $V$ increases. The lower bound of STOI and PSMt can be found for $\mu = 0$ as in AC and nSDP, and they decrease as $V$ increases. PSMt drops sharply for $V \geq 960$ and $\mu < 1$ particularly \rr{because} the interference is more dominant than the reproduced signal. Therefore, we can see this as AC and TIR are more important than nSDP in order to have higher STOI and/or PSMt. We can expect from the STOI metric that the reproduced sound is intelligible if $V \geq 120$ is used in this experiment\footnote{According to \cite{Taal2011}, a STOI score of more than $0.80$ maps to approximately $100$~\% speech intelligibility.}. On top of this, we can expect from PSMt that it has a better perception if $\mu \geq 1$. \mjr{We can expect that STOI and PSMt will not decrease in the case of one bright zone and one dark zone as $V$ increases or $\mu$ decreases because nSDP decreases.} A similar trend in these metrics is also observed from zone $\beta$. 
\input{chapters/04_simulations/exp3/exp3_summary_pcp.tex}

\subsection{Performance comparison}
\label{ssec:sim2_performComparison}
In the previous experiment, we investigated AP-VAST with respect to the physical and perceptual metrics as a function of $V$ and $\mu$ in Figs\ \ref{fig:exp1_results} and \ref{fig:exp2_results}, respectively. In this experiment, a comparison between AP-VAST and the reference methods is carried out. Specifically, how the signal-adaptive approach improves the performance is investigated by comparing VAST, P-VAST, and AP-VAST. The same input signals as in the previous experiment are used, and $\mu$ is set to $\mu = 1$. 

AC, nSDP, and TIR from the speech signal for zone $\alpha$ performed by five different methods are illustrated in Fig.~\ref{fig:exp3_results_py}. As seen in Fig.~\ref{fig:exp3_results_py}~\protect\subref{subfig:exp3_ac}, AC from all methods \rr{is} improved compared to the initial AC\rr{,} which is $0$~dB due to the symmetry of the system. ACC and PM provide around $15.7$~dB and $14.3$~dB of AC, respectively, whereas VAST and P-VAST vary depending on $V$\rr{,} but equal to or higher than $15$~dB. \rr{PM gives the lower bound of AC, and we can observe that AC by VAST converges to the lower bound as $V$ increases.} AP-VAST provides the highest AC across $V$ in this experiment and follows the same trend\rr{, which can be found in AC and nSDP,} as in the previous experiment.

As depicted in Fig.~\ref{fig:exp3_results_py}~\protect\subref{subfig:exp3_nsdp}, ACC and PM provide around $0$~dB and $-13.5$~dB of nSDP, respectively. \rr{Note that the initial nSDP is about $11.9$~dB}, but this is excluded in Fig.~\ref{fig:exp3_results_py}~\protect\subref{subfig:exp3_nsdp} for better visualization. Interestingly, nSDP of not only VAST but also P-VAST and AP-VAST seems to be upper- and lower-bounded by ACC and PM, respectively. By comparing Figs.~\ref{fig:exp3_results_py}~\protect\subref{subfig:exp3_ac} and \ref{fig:exp3_results_py}~\protect\subref{subfig:exp3_nsdp}, we can observe that higher AC yields larger nSDP from ACC, PM, and VAST\rr{,} which do not have perceptual weighting. However, this is not the case for P-VAST and AP-VAST since higher AC can be observed even at lower nSDP by comparing P-VAST and AP-VAST, e.g., when $V = 960$. As alluded to previously in this section, TIR follows the same trend in AC\rr{,} as seen in Figs.~\ref{fig:exp3_results_py}~\protect\subref{subfig:exp3_ac} and \protect\subref{subfig:exp3_tir}. 


In Fig.~\ref{fig:exp3_results_pp}, we observe STOI and PSMt. When \rr{there is} no control, a STOI of $0.6$ and a PSMt of $-0.01$ are observed but excluded in Fig.~\ref{fig:exp3_results_pp} for better visualization. We can observe that STOI and PSMt by VAST converge to that of PM as $V$ increases. P-VAST gives higher scores than VAST, and finally AP-VAST provides the highest STOI and PSMt amongst the methods across any $V$. 

From these experiments, we can observe that AP-VAST does not have the best performance across the physical metrics but does have it for the perceptual metrics.

\input{chapters/04_simulations/exp4/exp4_summary.tex}

\subsection{Performance comparison in a reverberant environment}
\label{ssec:sim3_performComparison_revb}
In the third and last experiment, we considered two input signals\rr{,} which were 4.5 seconds of track 49 (English female speech) and track 50 (English male speech) excerpt from the EBU SQAM database \cite{ebusqam2008}. \rr{They} were the desired signal in zone $\alpha$ and zone $\beta$, respectively. As in the signals used in the previous experiments, the energy of the signals was calibrated to be identical, and the signals were downsampled from $44.1$~kHz to $16$~kHz. The number of segments for AP-VAST was equal to $I = 135$.

For the reverberant environment, a room with $T_{60} = 200$~ms and \rr{a volume of} $140 \text{~m}^3$ was considered. In order to compare AP-VAST to the reference methods, the user parameters $V$ and $\mu$ are selected as $V = 1080$ and $\mu = 1$, respectively. Note that we can expect AP-VAST to have a lower nSDP \rr{as well as} higher AC, STOI, and PSMt if a high $\mu$ is selected\rr{,} which can be explained \rr{in} Figs.~\ref{fig:exp1_results} and \ref{fig:exp2_results}\rr{. However,} here $\mu = 1$ is specifically chosen in order to compare AP-VAST to PM \rr{directly}. Although the dereverberation might not be performed well \rr{because} the length of \rr{the} control filter is shorter than that of the reverberation, the performance comparison is still fair since this applies to all the candidate methods.

The performance of all the metrics as a function of mean and $95\%$ confidence interval is summarized in Table~\ref{tab:exp4_pfmmtx}. In general, compared to the performance in the anechoic environment depicted in Figs.~\ref{fig:exp3_results_py} and \ref{fig:exp3_results_pp} for $V = 1080$, performance degradation on all metrics is observed from all methods due to the reverberation. The highest AC, the minimum nSDP, and the largest TIR are obtained by P-VAST, PM, and VAST, respectively, but none of them provides the highest STOI or PSMt. Although AP-VAST provides neither the highest AC nor the lowest nSDP in this experiment, AP-VAST provides the highest STOI and PSMt. 

%% file: chapters/04_simulations/exp0/exp0_summary_ztpia_msk_info.tex
\begin{figure}[t]
	\centering
	\captionsetup[subfigure]{justification=centering}
		\subfloat[]{\includegraphics[scale=1]{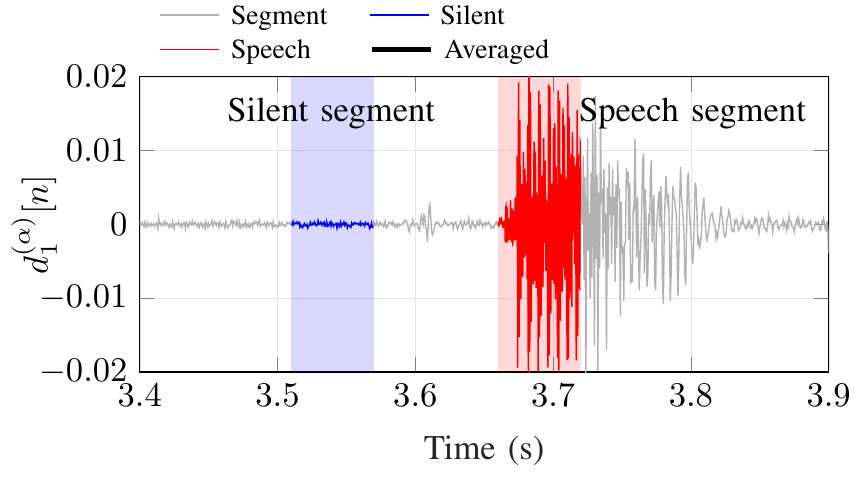}\label{subfig:exp0_signal}} 
		
		\subfloat[]{\includegraphics[scale=0.96]{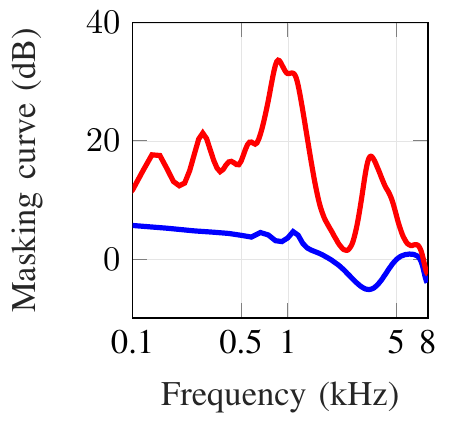}\label{subfig:exp0_sig_sil}}
		\hfill
		\subfloat[]{\includegraphics[scale=0.96]{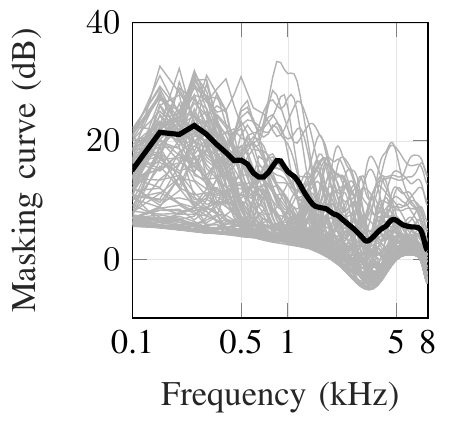}\label{subfig:exp0_allseg}}
		
		\caption{\protect\subref{subfig:exp0_signal} A subset of the speech signal for zone $\alpha$ in the first experiment, one of the speech segments and one of the silent segments are shown in red and in blue, respectively, \protect\subref{subfig:exp0_sig_sil} The masking curve (red) and the masking curve (blue) are computed from the speech segment and the silent segment, respectively, \protect\subref{subfig:exp0_allseg} Masking curves from each segment (gray) and the averaged masking curve across all the masking curves (black).}
		\label{fig:exp0_msk_info}
\end{figure}

%% file: chapters/04_simulations/exp1/exp1_summary.tex
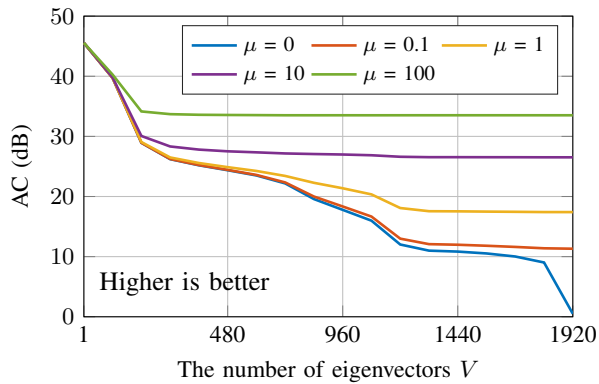
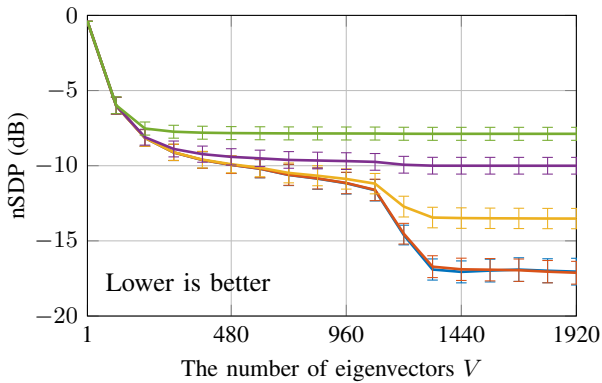
\begin{figure}[t]
	\centering
	\captionsetup[subfigure]{justification=centering}
		\subfloat[]{\input{chapters/04_simulations/exp1/exp1_apvast_ac_behavior.tex}\label{subfig:exp1_ac}}
		\quad 
		\subfloat[]{\input{chapters/04_simulations/exp1/exp1_apvast_nsdp_behavior.tex}\label{subfig:exp1_nsdp}}

		\caption{\cc{\protect\subref{subfig:exp1_ac} AC in dB and \protect\subref{subfig:exp1_nsdp} nSDP in dB as a function of $V$ for five different $\mu$: $\mu = 0$ (blue), $\mu = 0.1$ (red), $\mu = 1$ (yellow), $\mu = 10$ (purple), and $\mu = 100$ (green). Note that nSDP is represented with the $95\%$ confidence interval (error bars).}}
		\label{fig:exp1_results}
\end{figure}

%% file: chapters/04_simulations/exp1/exp1_apvast_ac_behavior.tex
%
%
\definecolor{mycolor1}{rgb}{0.00000,0.44700,0.74100}%
\definecolor{mycolor2}{rgb}{0.85000,0.32500,0.09800}%
\definecolor{mycolor3}{rgb}{0.92900,0.69400,0.12500}%
\definecolor{mycolor4}{rgb}{0.49400,0.18400,0.55600}%
\definecolor{mycolor5}{rgb}{0.46600,0.67400,0.18800}%
\begin{tikzpicture}

\begin{axis}[%
width = 6.5 cm,
height = 4.0 cm,
scale only axis,
xmin=1,
xmax=18,
xtick={1,6,10,14,18},
xticklabels={{1},{480},{960},{1440},{1920}},
ytick={0,10,20,30,40,50,60},
xlabel={The number of eigenvectors $V$},
ymin=0,
ymax=50,
ylabel={AC (dB)},
axis background/.style={fill=white},
grid = major,
legend style={legend cell align=left, align=left, draw=white!15!black, font=\footnotesize},
legend columns = 3, legend pos = north east
]
\addplot [color=mycolor1, line width=1pt]
  table[row sep=crcr]{%
1	45.5559623839593\\
2	39.7077064294856\\
3	28.9239108911263\\
4	26.2108130935314\\
5	25.1916578093612\\
6	24.3703993658787\\
7	23.5021877342356\\
8	22.1773344805484\\
9	19.5742627988282\\
10	17.7801487374901\\
11	15.9797037937574\\
12	12.017133802045\\
13	10.9973630871954\\
14	10.8401837415665\\
15	10.5310293874285\\
16	10.0211108519003\\
17	9.0266768076115\\
18	0.503857561521026\\
};
\addlegendentry{$\mu\text{ = 0}$}

\addplot [color=mycolor2, line width=1pt]
  table[row sep=crcr]{%
1	45.5559625205846\\
2	39.7083603825398\\
3	28.9377996367152\\
4	26.2403874698996\\
5	25.2316676076552\\
6	24.4259758481368\\
7	23.5885420626498\\
8	22.3393158190852\\
9	19.9950272373996\\
10	18.342416481977\\
11	16.6642336457685\\
12	12.9912921104741\\
13	12.0770699672597\\
14	11.978478327457\\
15	11.8109821672184\\
16	11.6159165669045\\
17	11.3877517756383\\
18	11.3174310884572\\
};
\addlegendentry{$\mu\text{ = 0.1}$}

\addplot [color=mycolor3, line width=1pt]
  table[row sep=crcr]{%
1	45.5559637501951\\
2	39.7142360856272\\
3	29.0599559839331\\
4	26.4950685894154\\
5	25.5702457142909\\
6	24.8802942923318\\
7	24.2484666358319\\
8	23.4208203311861\\
9	22.2773997761959\\
10	21.3704664331678\\
11	20.3521041077216\\
12	18.0839790239101\\
13	17.5552886577464\\
14	17.5342206622443\\
15	17.4891013897439\\
16	17.4545596490004\\
17	17.4115737117676\\
18	17.4066946785833\\
};
\addlegendentry{$\mu\text{ = 1}$}

\addplot [color=mycolor4, line width=1pt]
  table[row sep=crcr]{%
1	45.5559760445757\\
2	39.7720348848669\\
3	30.066772634052\\
4	28.3232669900581\\
5	27.8022775781801\\
6	27.5168773371932\\
7	27.346746342696\\
8	27.161548868216\\
9	27.0643357346419\\
10	26.9855379874128\\
11	26.8619622716544\\
12	26.6092289014495\\
13	26.5419552342574\\
14	26.5430453600216\\
15	26.5354201060191\\
16	26.5325620543225\\
17	26.5214864581664\\
18	26.5187809972698\\
};
\addlegendentry{$\mu\text{ = 10}$}

\addplot [color=mycolor5, line width=1pt]
  table[row sep=crcr]{%
1	45.5560988164725\\
2	40.2698011794084\\
3	34.1484741929199\\
4	33.6962168267742\\
5	33.592849575623\\
6	33.559655932841\\
7	33.537313253864\\
8	33.512258629857\\
9	33.5114054192469\\
10	33.512564434857\\
11	33.5096830734108\\
12	33.5129884580253\\
13	33.5139483578092\\
14	33.5139994462196\\
15	33.5132757284424\\
16	33.5136651603102\\
17	33.5121295778303\\
18	33.5114840303615\\
};
\addlegendentry{$\mu\text{ = 100}$}

\pgfplotsset{
	after end axis/.code={
		\node[black,above] at (axis cs:4.5, 2){Higher is better};
}}

\end{axis}
\end{tikzpicture}%

%% file: chapters/04_simulations/exp1/exp1_apvast_nsdp_behavior.tex
%
%
\definecolor{mycolor1}{rgb}{0.00000,0.44700,0.74100}%
\definecolor{mycolor2}{rgb}{0.85000,0.32500,0.09800}%
\definecolor{mycolor3}{rgb}{0.92900,0.69400,0.12500}%
\definecolor{mycolor4}{rgb}{0.49400,0.18400,0.55600}%
\definecolor{mycolor5}{rgb}{0.46600,0.67400,0.18800}%
\begin{tikzpicture}

\begin{axis}[%
width = 6.5 cm,
height = 4.0 cm,
scale only axis,
xmin=1,
xmax=18,
xtick={1,6,10,14,18},
xticklabels={{1},{480},{960},{1440},{1920}},
xlabel={The number of eigenvectors $V$},
ymin=-20,
ymax=0,
ylabel style={yshift=-1ex},
ylabel={nSDP (dB)},
axis background/.style={fill=white},
grid = major,
]
\addplot [color=mycolor1, line width=1pt]
 plot [error bars/.cd, y dir = both, y explicit]
 table[row sep=crcr, y error plus index=2, y error minus index=3]{%
1	-0.397314324969969	0.00647459589533113	0.00647459589533113\\
2	-6.00800660240516	0.562245160040967	0.562245160040967\\
3	-8.14988247304994	0.555950271857725	0.555950271857725\\
4	-9.11843399996974	0.537446208769674	0.537446208769674\\
5	-9.62045324806337	0.550696202416949	0.550696202416949\\
6	-9.94474318972575	0.572901105107041	0.572901105107041\\
7	-10.200323462563	0.605360847386668	0.605360847386668\\
8	-10.617238119674	0.704282543133276	0.704282543133276\\
9	-10.8589843529181	0.706726768565789	0.706726768565789\\
10	-11.1666165618709	0.716532203507269	0.716532203507269\\
11	-11.6305976314725	0.713757179417088	0.713757179417088\\
12	-14.6103801204977	0.654613813533058	0.654613813533058\\
13	-16.904773861022	0.703407606115899	0.703407606115899\\
14	-17.06284419634	0.729743989371382	0.729743989371382\\
15	-16.9759509517119	0.756655382193892	0.756655382193892\\
16	-16.9108972735874	0.786348779494286	0.786348779494286\\
17	-16.987420510883	0.797525870462774	0.797525870462774\\
18	-17.0474194955484	0.889723012962912	0.889723012962912\\
};

\addplot [color=mycolor2, line width=1pt]
 plot [error bars/.cd, y dir = both, y explicit]
 table[row sep=crcr, y error plus index=2, y error minus index=3]{%
1	-0.397314273970547	0.00647460489934814	0.00647460489934814\\
2	-6.00798283628889	0.562241143015361	0.562241143015361\\
3	-8.14973395341989	0.555642412820719	0.555642412820719\\
4	-9.11727293074084	0.537625628158974	0.537625628158974\\
5	-9.61842555976996	0.550874734264317	0.550874734264317\\
6	-9.94129309799122	0.572618263138873	0.572618263138873\\
7	-10.1953693204181	0.604251046595863	0.604251046595863\\
8	-10.6050962953839	0.698874486443918	0.698874486443918\\
9	-10.8465067343646	0.702793495098613	0.702793495098613\\
10	-11.149532559179	0.713325220964838	0.713325220964838\\
11	-11.6088633440303	0.713087035275914	0.713087035275914\\
12	-14.5225708754057	0.688162222157919	0.688162222157919\\
13	-16.7160733681134	0.727135743108063	0.727135743108063\\
14	-16.8922810463879	0.741249155204749	0.741249155204749\\
15	-16.9179401602122	0.743113252237429	0.743113252237429\\
16	-16.9517611427276	0.743738371537453	0.743738371537453\\
17	-17.0486513600169	0.743065118992683	0.743065118992683\\
18	-17.1194971051235	0.757459275178057	0.757459275178057\\
};

\addplot [color=mycolor3, line width=1pt]
 plot [error bars/.cd, y dir = both, y explicit]
 table[row sep=crcr, y error plus index=2, y error minus index=3]{%
1	-0.3973138148831	0.00647468593780868	0.00647468593780868\\
2	-6.00776828936805	0.562205023216374	0.562205023216374\\
3	-8.14796256683213	0.552878318870685	0.552878318870685\\
4	-9.10419129898184	0.538769663178877	0.538769663178877\\
5	-9.59475881999659	0.551902692723903	0.551902692723903\\
6	-9.90260524298852	0.569917922270547	0.569917922270547\\
7	-10.1371149153136	0.594808031382825	0.594808031382825\\
8	-10.4793531819479	0.660022592754724	0.660022592754724\\
9	-10.6668793418553	0.665837464750357	0.665837464750357\\
10	-10.8814483564199	0.674475152047733	0.674475152047733\\
11	-11.1931794127009	0.677359272223708	0.677359272223708\\
12	-12.7199319916219	0.690133835607398	0.690133835607398\\
13	-13.4477643885211	0.683745895255184	0.683745895255184\\
14	-13.4836431712787	0.682524693107117	0.682524693107117\\
15	-13.4976113291885	0.679085706350764	0.679085706350764\\
16	-13.5048776607271	0.675629099878115	0.675629099878115\\
17	-13.5138420871939	0.674639354769542	0.674639354769542\\
18	-13.5168645781343	0.67445406938161	0.67445406938161\\
};

\addplot [color=mycolor4, line width=1pt]
 plot [error bars/.cd, y dir = both, y explicit]
 table[row sep=crcr, y error plus index=2, y error minus index=3]{%
1	-0.397309214840972	0.00647549655158777	0.00647549655158777\\
2	-6.00556062745497	0.561846968701541	0.561846968701541\\
3	-8.10337063682104	0.526912939527751	0.526912939527751\\
4	-8.88542755314433	0.527396226481135	0.527396226481135\\
5	-9.23144009825671	0.537368103796503	0.537368103796503\\
6	-9.40974159402657	0.5409792335249	0.5409792335249\\
7	-9.51049794759184	0.545329365202215	0.545329365202215\\
8	-9.61514200065918	0.553352112187112	0.553352112187112\\
9	-9.65444906115152	0.554397311703347	0.554397311703347\\
10	-9.69325272014844	0.555844211805503	0.555844211805503\\
11	-9.74499980918972	0.556171733664788	0.556171733664788\\
12	-9.93490117472344	0.557967713394745	0.557967713394745\\
13	-9.99815377570684	0.55728426168228	0.55728426168228\\
14	-9.99984215663885	0.556897934535671	0.556897934535671\\
15	-10.0008734104157	0.556644821692052	0.556644821692052\\
16	-10.001291411109	0.556350307282906	0.556350307282906\\
17	-10.0018670007288	0.556346036811523	0.556346036811523\\
18	-10.0019901077503	0.556326391673051	0.556326391673051\\
};

\addplot [color=mycolor5, line width=1pt]
 plot [error bars/.cd, y dir = both, y explicit]
 table[row sep=crcr, y error plus index=2, y error minus index=3]{%
1	-0.397262302443322	0.00648362513891723	0.00648362513891723\\
2	-5.97904483055562	0.558456581183115	0.558456581183115\\
3	-7.52711800870554	0.435186698590436	0.435186698590436\\
4	-7.73836698589941	0.42886108456404	0.42886108456404\\
5	-7.80413551316115	0.430920900428701	0.430920900428701\\
6	-7.82988328903251	0.431253652872617	0.431253652872617\\
7	-7.84072329203045	0.431676053217424	0.431676053217424\\
8	-7.850361318095	0.432253977685228	0.432253977685228\\
9	-7.85352973864173	0.432327466294849	0.432327466294849\\
10	-7.85641851962343	0.432352054578827	0.432352054578827\\
11	-7.86024571551065	0.432352229059864	0.432352229059864\\
12	-7.87395016462734	0.432622359186188	0.432622359186188\\
13	-7.87832185696948	0.432717954892993	0.432717954892993\\
14	-7.87840277043568	0.432688113601299	0.432688113601299\\
15	-7.87847992038565	0.432683914333626	0.432683914333626\\
16	-7.87851447276936	0.432672802760575	0.432672802760575\\
17	-7.87855711310768	0.432677052180527	0.432677052180527\\
18	-7.87856853060544	0.432676575807441	0.432676575807441\\
};

\pgfplotsset{
	after end axis/.code={
		\node[black,above] at (axis cs:4.5, -19){Lower is better};
}}

\end{axis}
\end{tikzpicture}%

%% file: chapters/04_simulations/exp2/exp2_summary.tex
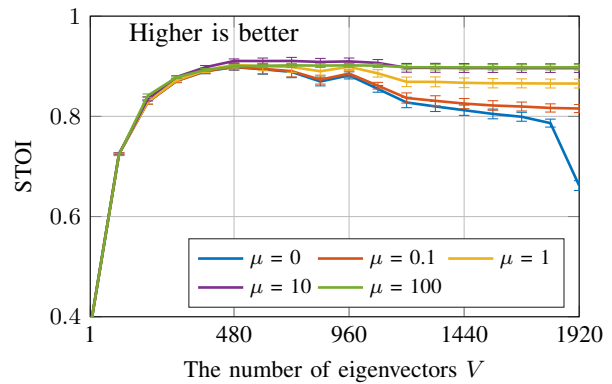
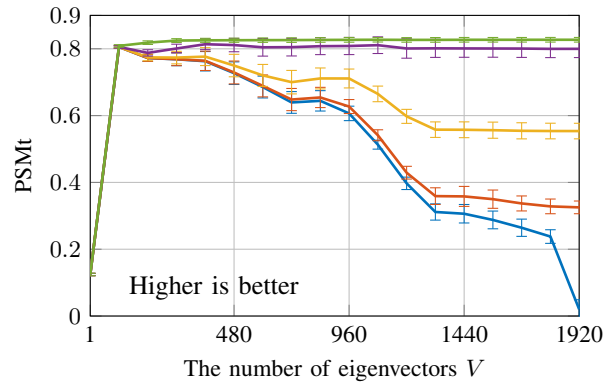
\begin{figure}[t]
	\centering
	\captionsetup[subfigure]{justification=centering}
		\subfloat[]{\input{chapters/04_simulations/exp2/exp2_apvast_stoi_behavior.tex}\label{subfig:exp2_stoi}}\quad
		\subfloat[]{\input{chapters/04_simulations/exp2/exp2_apvast_psmt_behavior.tex}\label{subfig:exp2_psmt}}
		
		\caption{\protect\subref{subfig:exp2_stoi} STOI in a range between $0.0$ and $1.0$ and \protect\subref{subfig:exp2_psmt} PSMt in a range of between $-1.0$ to $1.0$ as a function of $V$ for five different $\mu$: $\mu = 0$ (blue), $\mu = 0.1$ (red), $\mu = 1$ (yellow), $\mu = 10$ (purple), and $\mu = 100$ (green). Note that both are plotted with the $95\%$ confidence interval (error bars).}
		\label{fig:exp2_results}
\end{figure}

%% file: chapters/04_simulations/exp2/exp2_apvast_stoi_behavior.tex
%
%
\definecolor{mycolor1}{rgb}{0.00000,0.44700,0.74100}%
\definecolor{mycolor2}{rgb}{0.85000,0.32500,0.09800}%
\definecolor{mycolor3}{rgb}{0.92900,0.69400,0.12500}%
\definecolor{mycolor4}{rgb}{0.49400,0.18400,0.55600}%
\definecolor{mycolor5}{rgb}{0.46600,0.67400,0.18800}%
\begin{tikzpicture}

\begin{axis}[%
width = 6.5 cm,
height = 4.0 cm,
scale only axis,
xmin=1,
xmax=18,
xtick={1,6,10,14,18},
xticklabels={{1},{480},{960},{1440},{1920}},
xlabel={The number of eigenvectors $V$},
ymin=0.4,
ymax=1.0,
ylabel={STOI},
axis background/.style={fill=white},
grid = major,
legend style={legend cell align=left, align=left, draw=white!15!black, font=\footnotesize},
legend columns = 3, legend pos = south east
]
\addplot [color=mycolor1, line width=1pt]
 plot [error bars/.cd, y dir = both, y explicit]
 table[row sep=crcr, y error plus index=2, y error minus index=3]{%
1	0.377838322521512	0.00132141448576334	0.00132141448576334\\
2	0.72493459565598	0.0019677393312336	0.0019677393312336\\
3	0.828614943359208	0.00440403614024699	0.00440403614024699\\
4	0.870546067263013	0.00320516624159794	0.00320516624159794\\
5	0.890015860945251	0.00489513144031111	0.00489513144031111\\
6	0.898211700806573	0.00609882646969261	0.00609882646969261\\
7	0.89397160530879	0.00985366395680728	0.00985366395680728\\
8	0.888695930209652	0.0115138902803841	0.0115138902803841\\
9	0.869567463106543	0.00876333285729765	0.00876333285729765\\
10	0.881791418127468	0.00709227205334577	0.00709227205334577\\
11	0.85481356350188	0.00664906091065807	0.00664906091065807\\
12	0.827676148608688	0.010362136869783	0.010362136869783\\
13	0.819788415961933	0.01014143608631	0.01014143608631\\
14	0.812278310450025	0.0103164238543471	0.0103164238543471\\
15	0.804811306239258	0.00969016362698333	0.00969016362698333\\
16	0.799133069607008	0.00892281907136925	0.00892281907136925\\
17	0.786507981790592	0.00793248354829118	0.00793248354829118\\
18	0.661813710209437	0.00981292124119527	0.00981292124119527\\
};
\addlegendentry{$\mu\text{ = 0}$}

\addplot [color=mycolor2, line width=1pt]
 plot [error bars/.cd, y dir = both, y explicit]
 table[row sep=crcr, y error plus index=2, y error minus index=3]{%
1	0.37783831144773	0.00132141416017145	0.00132141416017145\\
2	0.724933753092868	0.00196792901169988	0.00196792901169988\\
3	0.828706642460145	0.00440065283065859	0.00440065283065859\\
4	0.870710654153076	0.00319028768319387	0.00319028768319387\\
5	0.890265576552862	0.004867037527381	0.004867037527381\\
6	0.898764560232142	0.00604598651856888	0.00604598651856888\\
7	0.894762088896085	0.0097532642056743	0.0097532642056743\\
8	0.890237907755951	0.0114314652451166	0.0114314652451166\\
9	0.873210647481906	0.00885073268353992	0.00885073268353992\\
10	0.885313221375611	0.00720757410865213	0.00720757410865213\\
11	0.861178260540942	0.00668273654139694	0.00668273654139694\\
12	0.836606759221172	0.0101602197103575	0.0101602197103575\\
13	0.830967745413546	0.0101316462285602	0.0101316462285602\\
14	0.82548999763346	0.0101367822381809	0.0101367822381809\\
15	0.821654987274049	0.00956149825970952	0.00956149825970952\\
16	0.819604931749415	0.00892967891526967	0.00892967891526967\\
17	0.816767715888007	0.00819329881320735	0.00819329881320735\\
18	0.815619305483589	0.00803283515710747	0.00803283515710747\\
};
\addlegendentry{$\mu\text{ = 0.1}$}

\addplot [color=mycolor3, line width=1pt]
 plot [error bars/.cd, y dir = both, y explicit]
 table[row sep=crcr, y error plus index=2, y error minus index=3]{%
1	0.377838211780999	0.00132141122966002	0.00132141122966002\\
2	0.724926122960478	0.00196963171781874	0.00196963171781874\\
3	0.829494559099311	0.004371579250388	0.004371579250388\\
4	0.872047410688049	0.00307123336575321	0.00307123336575321\\
5	0.892120694920595	0.00464115807470513	0.00464115807470513\\
6	0.902537157243761	0.0056497160332803	0.0056497160332803\\
7	0.899984524183061	0.00896741230345144	0.00896741230345144\\
8	0.898569808345837	0.0106959747027428	0.0106959747027428\\
9	0.890022426072223	0.0092467594148336	0.0092467594148336\\
10	0.898932913865746	0.00801443293992053	0.00801443293992053\\
11	0.885779158924514	0.00737576261169306	0.00737576261169306\\
12	0.868883175810267	0.00954595831167901	0.00954595831167901\\
13	0.868844549867564	0.00986643385653544	0.00986643385653544\\
14	0.86693433612473	0.00965859198109819	0.00965859198109819\\
15	0.866083301832936	0.00923928373382166	0.00923928373382166\\
16	0.865856604096727	0.00898017031591722	0.00898017031591722\\
17	0.865482625114187	0.00887538495797278	0.00887538495797278\\
18	0.865429091040952	0.0088433092051016	0.0088433092051016\\
};
\addlegendentry{$\mu\text{ = 1}$}

\addplot [color=mycolor4, line width=1pt]
 plot [error bars/.cd, y dir = both, y explicit]
 table[row sep=crcr, y error plus index=2, y error minus index=3]{%
1	0.377837214828846	0.00132138192378345	0.00132138192378345\\
2	0.724845224400252	0.00198632485237494	0.00198632485237494\\
3	0.834848875666134	0.00417190900574949	0.00417190900574949\\
4	0.878561527568486	0.00255497754242487	0.00255497754242487\\
5	0.89795050184304	0.00351191387143952	0.00351191387143952\\
6	0.910525353417467	0.00410246434377624	0.00410246434377624\\
7	0.910415244250498	0.00594122069218052	0.00594122069218052\\
8	0.910559729375857	0.00710381656311616	0.00710381656311616\\
9	0.908569116551878	0.00701881867897231	0.00701881867897231\\
10	0.909655875121051	0.00676650599033114	0.00676650599033114\\
11	0.906669071719004	0.00671699289830607	0.00671699289830607\\
12	0.896838449101737	0.0087469023146862	0.0087469023146862\\
13	0.896823950230584	0.0089241949869353	0.0089241949869353\\
14	0.896486359201533	0.00882782873651331	0.00882782873651331\\
15	0.896290328625175	0.00865359515613637	0.00865359515613637\\
16	0.896245975738222	0.00860493978874273	0.00860493978874273\\
17	0.896232074413737	0.00860330903107337	0.00860330903107337\\
18	0.896239754627449	0.0086026215672174	0.0086026215672174\\
};
\addlegendentry{$\mu\text{ = 10}$}

\addplot [color=mycolor5, line width=1pt]
 plot [error bars/.cd, y dir = both, y explicit]
 table[row sep=crcr, y error plus index=2, y error minus index=3]{%
1	0.377827223568593	0.0013210894593805	0.0013210894593805\\
2	0.7237325477035	0.00212590084789112	0.00212590084789112\\
3	0.841264550755252	0.00344557973991145	0.00344557973991145\\
4	0.877983668064419	0.0022077094832019	0.0022077094832019\\
5	0.892544913640006	0.00259731564564188	0.00259731564564188\\
6	0.900037308989862	0.00289997833548833	0.00289997833548833\\
7	0.900903624951614	0.00364915210607879	0.00364915210607879\\
8	0.901726163481708	0.0039023734847552	0.0039023734847552\\
9	0.90127186995756	0.00370050932879231	0.00370050932879231\\
10	0.901722451542151	0.00363227911773835	0.00363227911773835\\
11	0.901181425388661	0.0037074412567961	0.0037074412567961\\
12	0.898001827975687	0.00506284231912563	0.00506284231912563\\
13	0.897882267071254	0.00527461759692489	0.00527461759692489\\
14	0.897821689224381	0.00526778198605513	0.00526778198605513\\
15	0.897763754192054	0.00523011375821712	0.00523011375821712\\
16	0.897747875204764	0.00521316932433349	0.00521316932433349\\
17	0.897747102736897	0.00521069030961736	0.00521069030961736\\
18	0.897751319126931	0.00520952021131883	0.00520952021131883\\
};
\addlegendentry{$\mu\text{ = 100}$}

\pgfplotsset{
	after end axis/.code={
		\node[black,above] at (axis cs:5.3, 0.92){Higher is better};
}}

\end{axis}
\end{tikzpicture}%

%% file: chapters/04_simulations/exp2/exp2_apvast_psmt_behavior.tex
%
%
\definecolor{mycolor1}{rgb}{0.00000,0.44700,0.74100}%
\definecolor{mycolor2}{rgb}{0.85000,0.32500,0.09800}%
\definecolor{mycolor3}{rgb}{0.92900,0.69400,0.12500}%
\definecolor{mycolor4}{rgb}{0.49400,0.18400,0.55600}%
\definecolor{mycolor5}{rgb}{0.46600,0.67400,0.18800}%
\begin{tikzpicture}

\begin{axis}[%
width = 6.5 cm,
height = 4.0 cm,
scale only axis,
xmin=1,
xmax=18,
xtick={1,6,10,14,18},
xticklabels={{1},{480},{960},{1440},{1920}},
ytick={0.0, 0.2, 0.4, 0.6, 0.8, 0.9},
xlabel = {The number of eigenvectors $V$},
ymin = 0.0,
ymax = 0.9,
ylabel={PSMt},
axis background/.style={fill=white},
grid = major,
]

\addplot [color=mycolor1, line width=1pt]
 plot [error bars/.cd, y dir = both, y explicit]
 table[row sep=crcr, y error plus index=2, y error minus index=3]{%
1	0.124315755658928	0.00350548527126188	0.00350548527126188\\
2	0.805485704494445	0.0036807377628757	0.0036807377628757\\
3	0.77229132412569	0.010115292729492	0.010115292729492\\
4	0.768704462582237	0.0193417098338368	0.0193417098338368\\
5	0.762755415197588	0.0290288019539634	0.0290288019539634\\
6	0.727027446343388	0.0338454703608479	0.0338454703608479\\
7	0.685528574872275	0.03265242472782	0.03265242472782\\
8	0.639395872174953	0.0324786624133465	0.0324786624133465\\
9	0.64420214868442	0.0304940583628783	0.0304940583628783\\
10	0.606607655878095	0.0216512717879758	0.0216512717879758\\
11	0.513030306176987	0.0134024670867116	0.0134024670867116\\
12	0.397233489729287	0.0185754681940398	0.0185754681940398\\
13	0.311464592916513	0.0243073500669697	0.0243073500669697\\
14	0.306230055237321	0.027655725188882	0.027655725188882\\
15	0.287692975905903	0.0265506426644277	0.0265506426644277\\
16	0.264307739495252	0.0258105285351937	0.0258105285351937\\
17	0.237928193262168	0.0203900122747686	0.0203900122747686\\
18	0.0200010556879685	0.0287232590529206	0.0287232590529206\\
};

\addplot [color=mycolor2, line width=1pt]
 plot [error bars/.cd, y dir = both, y explicit]
 table[row sep=crcr, y error plus index=2, y error minus index=3]{%
1	0.124315779127257	0.00350556254841961	0.00350556254841961\\
2	0.80549621086362	0.00368285850405823	0.00368285850405823\\
3	0.772453536928663	0.0101301120693523	0.0101301120693523\\
4	0.768903994430461	0.0192982249698842	0.0192982249698842\\
5	0.76394101312783	0.0288902020745182	0.0288902020745182\\
6	0.729133054781612	0.0342417829673288	0.0342417829673288\\
7	0.688994592843701	0.0328872405017395	0.0328872405017395\\
8	0.648075846665823	0.033222386012011	0.033222386012011\\
9	0.654314123610922	0.0303122329070308	0.0303122329070308\\
10	0.627135366196839	0.0209222786999207	0.0209222786999207\\
11	0.541193352227509	0.0163720366357394	0.0163720366357394\\
12	0.429131643295742	0.0190845797983695	0.0190845797983695\\
13	0.358963455394582	0.0249325766290846	0.0249325766290846\\
14	0.358107164481055	0.0299200786250978	0.0299200786250978\\
15	0.349832540189857	0.0273638434128521	0.0273638434128521\\
16	0.337046375935733	0.022439119351321	0.022439119351321\\
17	0.328230936947986	0.022003818801063	0.022003818801063\\
18	0.325288958991045	0.0190143609457103	0.0190143609457103\\
};

\addplot [color=mycolor3, line width=1pt]
 plot [error bars/.cd, y dir = both, y explicit]
 table[row sep=crcr, y error plus index=2, y error minus index=3]{%
1	0.124315752990408	0.00350587099602675	0.00350587099602675\\
2	0.805526223781089	0.00368958432510378	0.00368958432510378\\
3	0.7743426209482	0.0101796919798208	0.0101796919798208\\
4	0.773670051557112	0.0187949440374031	0.0187949440374031\\
5	0.776588414825642	0.0268125174974928	0.0268125174974928\\
6	0.749850919918626	0.0336490148519341	0.0336490148519341\\
7	0.720357080437978	0.0328830137067151	0.0328830137067151\\
8	0.700143548005496	0.0351846218325211	0.0351846218325211\\
9	0.711255387220316	0.0313605958673934	0.0313605958673934\\
10	0.711127008286766	0.0282868582925804	0.0282868582925804\\
11	0.664978326036003	0.0234033231389224	0.0234033231389224\\
12	0.597758215552322	0.0209609880514714	0.0209609880514714\\
13	0.557961102277973	0.0232108095105712	0.0232108095105712\\
14	0.557603416465537	0.0236716561472649	0.0236716561472649\\
15	0.556354434111545	0.024168201885433	0.024168201885433\\
16	0.554475713714365	0.0241282075235418	0.0241282075235418\\
17	0.553840370100754	0.0234068375907087	0.0234068375907087\\
18	0.553719222211201	0.0232900973965414	0.0232900973965414\\
};

\addplot [color=mycolor4, line width=1pt]
 plot [error bars/.cd, y dir = both, y explicit]
 table[row sep=crcr, y error plus index=2, y error minus index=3]{%
1	0.124315522146797	0.00350961223524814	0.00350961223524814\\
2	0.805574961718114	0.003706024254031	0.003706024254031\\
3	0.787735636570331	0.00948968741121456	0.00948968741121456\\
4	0.800922839061679	0.0129032962149201	0.0129032962149201\\
5	0.813670429692344	0.0162544580206438	0.0162544580206438\\
6	0.81119523880517	0.0200435848582384	0.0200435848582384\\
7	0.804472179025392	0.0263809116306897	0.0263809116306897\\
8	0.804895304049823	0.0263294559052724	0.0263294559052724\\
9	0.807781586642239	0.0252228474997737	0.0252228474997737\\
10	0.808375380690227	0.0256994311735148	0.0256994311735148\\
11	0.810828045661145	0.024959308498182	0.024959308498182\\
12	0.801001098120172	0.0291185670858132	0.0291185670858132\\
13	0.801403061967298	0.0271365141639634	0.0271365141639634\\
14	0.801099832200013	0.0264926436393301	0.0264926436393301\\
15	0.800873046994776	0.0260807676713908	0.0260807676713908\\
16	0.800673844656587	0.0264468827080953	0.0264468827080953\\
17	0.799799878072996	0.0261738249792152	0.0261738249792152\\
18	0.799883159015115	0.0261215603891908	0.0261215603891908\\
};

\addplot [color=mycolor5, line width=1pt]
 plot [error bars/.cd, y dir = both, y explicit]
 table[row sep=crcr, y error plus index=2, y error minus index=3]{%
1	0.12430767671016	0.00354416721085469	0.00354416721085469\\
2	0.808681388719487	0.00381262877020856	0.00381262877020856\\
3	0.818423467236576	0.00542202537277077	0.00542202537277077\\
4	0.824688560471861	0.0056967921428382	0.0056967921428382\\
5	0.825390257635768	0.00604842127941284	0.00604842127941284\\
6	0.825652228666476	0.00623356732664823	0.00623356732664823\\
7	0.825945006489004	0.00665343602483891	0.00665343602483891\\
8	0.826089683382826	0.0068730042784647	0.0068730042784647\\
9	0.826519965452032	0.00697543828960211	0.00697543828960211\\
10	0.826851779770734	0.00712580911319588	0.00712580911319588\\
11	0.82710334142412	0.00677534096809848	0.00677534096809848\\
12	0.826675618963431	0.00689216105924874	0.00689216105924874\\
13	0.8267139553765	0.00703569487764318	0.00703569487764318\\
14	0.826991059389206	0.00695974446852019	0.00695974446852019\\
15	0.827010615249222	0.00693305493941682	0.00693305493941682\\
16	0.827006739964714	0.00692533141210506	0.00692533141210506\\
17	0.827059418202097	0.00698138102249374	0.00698138102249374\\
18	0.827043641219546	0.00698886906898896	0.00698886906898896\\
};

\pgfplotsset{
	after end axis/.code={
		\node[black,above] at (axis cs:5.3, 0.02){Higher is better};
}}

\end{axis}
\end{tikzpicture}%

%% file: chapters/04_simulations/exp3/exp3_summary_phy.tex
\begin{figure}[t]
	\centering
	\captionsetup[subfigure]{justification=centering}
		\subfloat[]{\input{chapters/04_simulations/exp3/exp3_all_free_ac.tex}\label{subfig:exp3_ac}}
		\quad
		\subfloat[]{\input{chapters/04_simulations/exp3/exp3_all_free_nsdp.tex}\label{subfig:exp3_nsdp}}
		\quad
		\subfloat[]{\input{chapters/04_simulations/exp3/exp3_all_free_tir.tex}\label{subfig:exp3_tir}}
		
		\caption{\cc{\protect\subref{subfig:exp3_ac} AC in dB, \protect\subref{subfig:exp3_nsdp} nSDP in dB, and \protect\subref{subfig:exp3_tir} TIR in dB as a function of $V$ for five different methods: PM (purple dash-dot) \cite{Simon-Galvez2015}, ACC (black dash-dot) \cite{choi2002generation}, VAST (blue dash) \cite{Lee2018}, P-VAST (green dot), and AP-VAST (red solid). Note that nSDP and TIR are represented with the $95\%$ confidence interval (error bars).}}
		\label{fig:exp3_results_py}
\end{figure}
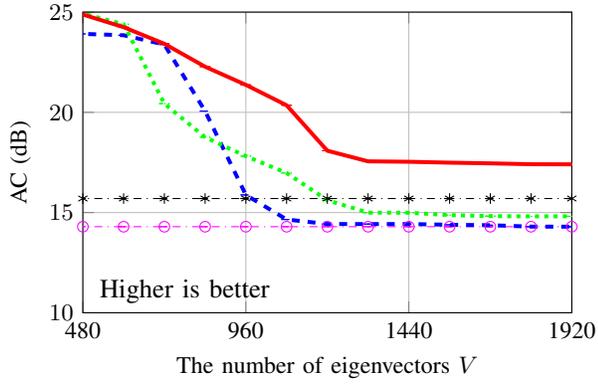
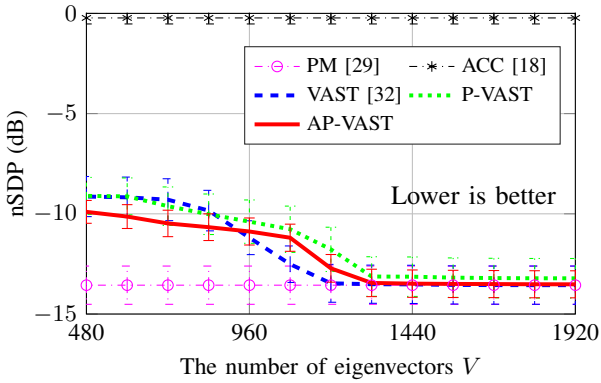
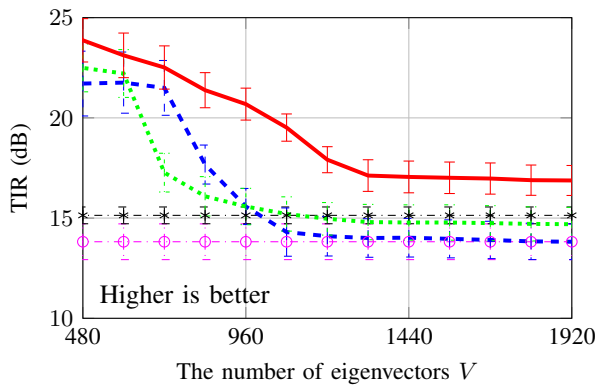

%% file: chapters/04_simulations/exp3/exp3_all_free_ac.tex
%
%
\definecolor{mycolor2}{rgb}{0.00000,1.00000,1.00000}%
\definecolor{mycolor1}{rgb}{1.00000,0.00000,1.00000}%
\begin{tikzpicture}

\begin{axis}[%
width = 6.5 cm,
height = 4.0 cm,
scale only axis,
xmin=6,
xmax=18,
xtick={1,6,10,14,18},
xticklabels={{1},{480},{960},{1440},{1920}},
xlabel={The number of eigenvectors $V$},
ymin=10,
ymax=25,
ylabel={AC (dB)},
axis background/.style={fill=white},
grid = major,
legend style={legend cell align=left, align=left, draw=white!15!black, font=\footnotesize},
legend columns = 2, legend pos = north east
]
\addplot [color=mycolor1, dashdotted, mark=o, mark options={solid, mycolor1}]
 plot [error bars/.cd, y dir = both, y explicit]
 table[row sep=crcr, y error plus index=2, y error minus index=3]{%
1	14.2945477677033	0	0\\
2	14.2945477677033	0	0\\
3	14.2945477677033	0	0\\
4	14.2945477677033	0	0\\
5	14.2945477677033	0	0\\
6	14.2945477677033	0	0\\
7	14.2945477677033	0	0\\
8	14.2945477677033	0	0\\
9	14.2945477677033	0	0\\
10	14.2945477677033	0	0\\
11	14.2945477677033	0	0\\
12	14.2945477677033	0	0\\
13	14.2945477677033	0	0\\
14	14.2945477677033	0	0\\
15	14.2945477677033	0	0\\
16	14.2945477677033	0	0\\
17	14.2945477677033	0	0\\
18	14.2945477677033	0	0\\
};


\addplot [color=black, dashdotted, mark=asterisk, mark options={solid, black}]
 plot [error bars/.cd, y dir = both, y explicit]
 table[row sep=crcr, y error plus index=2, y error minus index=3]{%
1	15.7003406969929	0	0\\
2	15.7003406969929	0	0\\
3	15.7003406969929	0	0\\
4	15.7003406969929	0	0\\
5	15.7003406969929	0	0\\
6	15.7003406969929	0	0\\
7	15.7003406969929	0	0\\
8	15.7003406969929	0	0\\
9	15.7003406969929	0	0\\
10	15.7003406969929	0	0\\
11	15.7003406969929	0	0\\
12	15.7003406969929	0	0\\
13	15.7003406969929	0	0\\
14	15.7003406969929	0	0\\
15	15.7003406969929	0	0\\
16	15.7003406969929	0	0\\
17	15.7003406969929	0	0\\
18	15.7003406969929	0	0\\
};


\addplot [color=blue, dashed, line width=1.5pt]
 plot [error bars/.cd, y dir = both, y explicit]
 table[row sep=crcr, y error plus index=2, y error minus index=3]{%
1	46.5800387524906	0	0\\
2	40.4036683822506	0	0\\
3	28.8016288432132	0	0\\
4	26.2462420895804	0	0\\
5	26.1800272158339	0	0\\
6	23.9144072754278	0	0\\
7	23.8470236210559	0	0\\
8	23.3814651016721	0	0\\
9	20.0635220289225	0	0\\
10	15.8793067118469	0	0\\
11	14.6544243526078	0	0\\
12	14.4298117426931	0	0\\
13	14.4253360827768	0	0\\
14	14.4280158158853	0	0\\
15	14.390549564875	0	0\\
16	14.3685782079648	0	0\\
17	14.298415603314	0	0\\
18	14.2945477677033	0	0\\
};

\addplot [color=green, dotted, line width=1.5pt]
 plot [error bars/.cd, y dir = both, y explicit]
 table[row sep=crcr, y error plus index=2, y error minus index=3]{%
1	46.9441675621935	0	0\\
2	40.1330472124511	0	0\\
3	28.7141899639102	0	0\\
4	26.4935660384716	0	0\\
5	25.6775860317476	0	0\\
6	24.92628680091	0	0\\
7	24.396353460908	0	0\\
8	20.4352946877973	0	0\\
9	18.7572702584826	0	0\\
10	17.8153646405637	0	0\\
11	16.9742326203764	0	0\\
12	15.6157451995988	0	0\\
13	14.9966620486475	0	0\\
14	14.9893846939181	0	0\\
15	14.8737237516181	0	0\\
16	14.8219441055181	0	0\\
17	14.8100474488242	0	0\\
18	14.8135382311419	0	0\\
};

\addplot [color=red, line width=1.5pt]
 plot [error bars/.cd, y dir = both, y explicit]
 table[row sep=crcr, y error plus index=2, y error minus index=3]{%
1	45.5559637501951	0	0\\
2	39.7142360856272	0	0\\
3	29.0599559839331	0	0\\
4	26.4950685894154	0	0\\
5	25.5702457142909	0	0\\
6	24.8802942923318	0	0\\
7	24.2484666358319	0	0\\
8	23.4208203311861	0	0\\
9	22.2773997761959	0	0\\
10	21.3704664331678	0	0\\
11	20.3521041077216	0	0\\
12	18.0839790239101	0	0\\
13	17.5552886577464	0	0\\
14	17.5342206622443	0	0\\
15	17.4891013897439	0	0\\
16	17.4545596490004	0	0\\
17	17.4115737117676	0	0\\
18	17.4066946785833	0	0\\
};

\pgfplotsset{
	after end axis/.code={
		\node[black,above] at (axis cs:8.5, 10){Higher is better};
}}

\end{axis}

\end{tikzpicture}%

%% file: chapters/04_simulations/exp3/exp3_all_free_nsdp.tex
%
%
\definecolor{mycolor2}{rgb}{0.00000,1.00000,1.00000}%
\definecolor{mycolor1}{rgb}{1.00000,0.00000,1.00000}%
\begin{tikzpicture}

\begin{axis}[%
width = 6.5 cm,
height = 4.0 cm,
scale only axis,
xmin=6,
xmax=18,
xtick={1,6,10,14,18},
xticklabels={{1},{480},{960},{1440},{1920}},
xlabel={The number of eigenvectors $V$},
ymin=-15,
ymax=0,
ylabel style={yshift=-1ex},
ylabel={nSDP (dB)},
axis background/.style={fill=white},
grid = major,
legend style={legend cell align=left, align=left, draw=white!15!black, font=\footnotesize, yshift=-2ex},
legend columns = 2, legend pos = north east
]

\addplot [color=mycolor1, dashdotted, mark=o, mark options={solid, mycolor1}]
 plot [error bars/.cd, y dir = both, y explicit]
 table[row sep=crcr, y error plus index=2, y error minus index=3]{%
1	-13.5601752205251	0.954191748291462	0.954191748291462\\
2	-13.5601752205251	0.954191748291462	0.954191748291462\\
3	-13.5601752205251	0.954191748291462	0.954191748291462\\
4	-13.5601752205251	0.954191748291462	0.954191748291462\\
5	-13.5601752205251	0.954191748291462	0.954191748291462\\
6	-13.5601752205251	0.954191748291462	0.954191748291462\\
7	-13.5601752205251	0.954191748291462	0.954191748291462\\
8	-13.5601752205251	0.954191748291462	0.954191748291462\\
9	-13.5601752205251	0.954191748291462	0.954191748291462\\
10	-13.5601752205251	0.954191748291462	0.954191748291462\\
11	-13.5601752205251	0.954191748291462	0.954191748291462\\
12	-13.5601752205251	0.954191748291462	0.954191748291462\\
13	-13.5601752205251	0.954191748291462	0.954191748291462\\
14	-13.5601752205251	0.954191748291462	0.954191748291462\\
15	-13.5601752205251	0.954191748291462	0.954191748291462\\
16	-13.5601752205251	0.954191748291462	0.954191748291462\\
17	-13.5601752205251	0.954191748291462	0.954191748291462\\
18	-13.5601752205251	0.954191748291462	0.954191748291462\\
};
\addlegendentry{PM \cite{Simon-Galvez2015}}


\addplot [color=black, dashdotted, mark=asterisk, mark options={solid, black}]
 plot [error bars/.cd, y dir = both, y explicit]
 table[row sep=crcr, y error plus index=2, y error minus index=3]{%
1	-0.226315552599528	0.300369194658847	0.300369194658847\\
2	-0.226315552599528	0.300369194658847	0.300369194658847\\
3	-0.226315552599528	0.300369194658847	0.300369194658847\\
4	-0.226315552599528	0.300369194658847	0.300369194658847\\
5	-0.226315552599528	0.300369194658847	0.300369194658847\\
6	-0.226315552599528	0.300369194658847	0.300369194658847\\
7	-0.226315552599528	0.300369194658847	0.300369194658847\\
8	-0.226315552599528	0.300369194658847	0.300369194658847\\
9	-0.226315552599528	0.300369194658847	0.300369194658847\\
10	-0.226315552599528	0.300369194658847	0.300369194658847\\
11	-0.226315552599528	0.300369194658847	0.300369194658847\\
12	-0.226315552599528	0.300369194658847	0.300369194658847\\
13	-0.226315552599528	0.300369194658847	0.300369194658847\\
14	-0.226315552599528	0.300369194658847	0.300369194658847\\
15	-0.226315552599528	0.300369194658847	0.300369194658847\\
16	-0.226315552599528	0.300369194658847	0.300369194658847\\
17	-0.226315552599528	0.300369194658847	0.300369194658847\\
18	-0.226315552599528	0.300369194658847	0.300369194658847\\
};
\addlegendentry{ACC \cite{choi2002generation}}


\addplot [color=blue, dashed, line width=1.5pt]
 plot [error bars/.cd, y dir = both, y explicit]
 table[row sep=crcr, y error plus index=2, y error minus index=3]{%
1	-0.423144834432958	0.00907741559274575	0.00907741559274575\\
2	-5.830056061021	0.724025810257694	0.724025810257694\\
3	-7.89671878772689	0.832014690148184	0.832014690148184\\
4	-8.67093605091206	0.913937855460517	0.913937855460517\\
5	-8.81715874525597	0.929653015937812	0.929653015937812\\
6	-9.13833092381344	1.00053358868198	1.00053358868198\\
7	-9.17696998550108	1.01016577725642	1.01016577725642\\
8	-9.29522082515144	1.04497254101476	1.04497254101476\\
9	-9.83906100227889	1.01090441336538	1.01090441336538\\
10	-11.1909788064458	0.844303118639221	0.844303118639221\\
11	-12.5137519485484	0.911209346200548	0.911209346200548\\
12	-13.467452842291	0.944717707887413	0.944717707887413\\
13	-13.511401019311	0.952817058240332	0.952817058240332\\
14	-13.5205333473353	0.953762553158887	0.953762553158887\\
15	-13.5480572081219	0.95201815052822	0.95201815052822\\
16	-13.5559884974557	0.951997470572188	0.951997470572188\\
17	-13.5595696913474	0.95425381346815	0.95425381346815\\
18	-13.5601752205251	0.954191748291462	0.954191748291462\\
};
\addlegendentry{VAST \cite{Lee2018}}

\addplot [color=green, dotted, line width=1.5pt]
 plot [error bars/.cd, y dir = both, y explicit]
 table[row sep=crcr, y error plus index=2, y error minus index=3]{%
1	-0.272857553001428	0.00402939564642891	0.00402939564642891\\
2	-5.9905956289095	0.656935315361609	0.656935315361609\\
3	-8.00637001643435	0.666523029470864	0.666523029470864\\
4	-8.57267546869059	0.789105817939374	0.789105817939374\\
5	-8.8758466044384	0.909789598685322	0.909789598685322\\
6	-9.10534749617691	0.920823374231127	0.920823374231127\\
7	-9.12978462464121	0.920539118883155	0.920539118883155\\
8	-9.61952029794989	0.961020464176272	0.961020464176272\\
9	-10.0218123082144	1.01107429363658	1.01107429363658\\
10	-10.3895465627377	1.08158773605448	1.08158773605448\\
11	-10.7697144321149	1.15489330777146	1.15489330777146\\
12	-11.773011065164	1.1047403851301	1.1047403851301\\
13	-13.1285398589911	0.987354187300861	0.987354187300861\\
14	-13.1366004661485	0.984323234883059	0.984323234883059\\
15	-13.1841550360358	0.984268026569722	0.984268026569722\\
16	-13.2053359996603	0.987097831146838	0.987097831146838\\
17	-13.2156627532464	0.988073970316809	0.988073970316809\\
18	-13.217727158966	0.988717668519548	0.988717668519548\\
};
\addlegendentry{P-VAST}

\addplot [color=red, line width=1.5pt]
 plot [error bars/.cd, y dir = both, y explicit]
 table[row sep=crcr, y error plus index=2, y error minus index=3]{%
1	-0.3973138148831	0.00647468593780868	0.00647468593780868\\
2	-6.00776828936805	0.562205023216374	0.562205023216374\\
3	-8.14796256683213	0.552878318870685	0.552878318870685\\
4	-9.10419129898184	0.538769663178877	0.538769663178877\\
5	-9.59475881999659	0.551902692723903	0.551902692723903\\
6	-9.90260524298852	0.569917922270547	0.569917922270547\\
7	-10.1371149153136	0.594808031382825	0.594808031382825\\
8	-10.4793531819479	0.660022592754724	0.660022592754724\\
9	-10.6668793418553	0.665837464750357	0.665837464750357\\
10	-10.8814483564199	0.674475152047733	0.674475152047733\\
11	-11.1931794127009	0.677359272223708	0.677359272223708\\
12	-12.7199319916219	0.690133835607398	0.690133835607398\\
13	-13.4477643885211	0.683745895255184	0.683745895255184\\
14	-13.4836431712787	0.682524693107117	0.682524693107117\\
15	-13.4976113291885	0.679085706350764	0.679085706350764\\
16	-13.5048776607271	0.675629099878115	0.675629099878115\\
17	-13.5138420871939	0.674639354769542	0.674639354769542\\
18	-13.5168645781343	0.67445406938161	0.67445406938161\\
};
\addlegendentry{AP-VAST}

\pgfplotsset{
	after end axis/.code={
		\node[black,above] at (axis cs:15.5, -10.0){Lower is better};
}}

\end{axis}
\end{tikzpicture}%

%% file: chapters/04_simulations/exp3/exp3_all_free_tir.tex
%
%
\definecolor{mycolor2}{rgb}{0.00000,1.00000,1.00000}%
\definecolor{mycolor1}{rgb}{1.00000,0.00000,1.00000}%
\begin{tikzpicture}

\begin{axis}[%
width = 6.5 cm,
height = 4.0 cm,
scale only axis,
xmin=6,
xmax=18,
xtick={1,6,10,14,18},
xticklabels={{1},{480},{960},{1440},{1920}},
xlabel={The number of eigenvectors $V$},
ymin=10,
ymax=25,
ylabel={TIR (dB)},
axis background/.style={fill=white},
grid = major,
legend style={legend cell align=left, align=left, draw=white!15!black, font=\footnotesize},
legend columns = 2, legend pos = north east
]
\addplot [color=mycolor1, dashdotted, mark=o, mark options={solid, mycolor1}]
 plot [error bars/.cd, y dir = both, y explicit]
 table[row sep=crcr, y error plus index=2, y error minus index=3]{%
1	13.8198143012847	0.892008570860182	0.892008570860182\\
2	13.8198143012847	0.892008570860182	0.892008570860182\\
3	13.8198143012847	0.892008570860182	0.892008570860182\\
4	13.8198143012847	0.892008570860182	0.892008570860182\\
5	13.8198143012847	0.892008570860182	0.892008570860182\\
6	13.8198143012847	0.892008570860182	0.892008570860182\\
7	13.8198143012847	0.892008570860182	0.892008570860182\\
8	13.8198143012847	0.892008570860182	0.892008570860182\\
9	13.8198143012847	0.892008570860182	0.892008570860182\\
10	13.8198143012847	0.892008570860182	0.892008570860182\\
11	13.8198143012847	0.892008570860182	0.892008570860182\\
12	13.8198143012847	0.892008570860182	0.892008570860182\\
13	13.8198143012847	0.892008570860182	0.892008570860182\\
14	13.8198143012847	0.892008570860182	0.892008570860182\\
15	13.8198143012847	0.892008570860182	0.892008570860182\\
16	13.8198143012847	0.892008570860182	0.892008570860182\\
17	13.8198143012847	0.892008570860182	0.892008570860182\\
18	13.8198143012847	0.892008570860182	0.892008570860182\\
};


\addplot [color=black, dashdotted, mark=asterisk, mark options={solid, black}]
 plot [error bars/.cd, y dir = both, y explicit]
 table[row sep=crcr, y error plus index=2, y error minus index=3]{%
1	15.1364674682868	0.418555408081725	0.418555408081725\\
2	15.1364674682868	0.418555408081725	0.418555408081725\\
3	15.1364674682868	0.418555408081725	0.418555408081725\\
4	15.1364674682868	0.418555408081725	0.418555408081725\\
5	15.1364674682868	0.418555408081725	0.418555408081725\\
6	15.1364674682868	0.418555408081725	0.418555408081725\\
7	15.1364674682868	0.418555408081725	0.418555408081725\\
8	15.1364674682868	0.418555408081725	0.418555408081725\\
9	15.1364674682868	0.418555408081725	0.418555408081725\\
10	15.1364674682868	0.418555408081725	0.418555408081725\\
11	15.1364674682868	0.418555408081725	0.418555408081725\\
12	15.1364674682868	0.418555408081725	0.418555408081725\\
13	15.1364674682868	0.418555408081725	0.418555408081725\\
14	15.1364674682868	0.418555408081725	0.418555408081725\\
15	15.1364674682868	0.418555408081725	0.418555408081725\\
16	15.1364674682868	0.418555408081725	0.418555408081725\\
17	15.1364674682868	0.418555408081725	0.418555408081725\\
18	15.1364674682868	0.418555408081725	0.418555408081725\\
};


\addplot [color=blue, dashed, line width=1.5pt]
 plot [error bars/.cd, y dir = both, y explicit]
 table[row sep=crcr, y error plus index=2, y error minus index=3]{%
1	47.8769007136653	2.05533957742191	2.05533957742191\\
2	40.9365498242196	1.85547011360792	1.85547011360792\\
3	27.5657430194762	1.15431367675851	1.15431367675851\\
4	23.3883242330205	1.41444977473396	1.41444977473396\\
5	23.3136025218884	1.49646364110653	1.49646364110653\\
6	21.7114411012182	1.61661213892322	1.61661213892322\\
7	21.7591930569874	1.52571444146349	1.52571444146349\\
8	21.4957607245787	1.36733603847907	1.36733603847907\\
9	17.6712157438107	0.972883683854164	0.972883683854164\\
10	15.5781191811356	0.899186235828223	0.899186235828223\\
11	14.3016417843846	1.20505284321353	1.20505284321353\\
12	14.0954074206602	0.986420092280879	0.986420092280879\\
13	14.0060079701283	0.961820590855216	0.961820590855216\\
14	14.0212505673972	0.96451420766206	0.96451420766206\\
15	13.9640745926874	0.947742898527943	0.947742898527943\\
16	13.9095774215607	0.931460019784133	0.931460019784133\\
17	13.8375647866821	0.908365599827765	0.908365599827765\\
18	13.8198143012847	0.892008570860182	0.892008570860182\\
};

\addplot [color=green, dotted, line width=1.5pt]
 plot [error bars/.cd, y dir = both, y explicit]
 table[row sep=crcr, y error plus index=2, y error minus index=3]{%
1	45.5501232103496	1.4520874683185	1.4520874683185\\
2	41.171593663492	1.46068307126449	1.46068307126449\\
3	28.8549020161089	1.0436574958967	1.0436574958967\\
4	27.5500993521852	0.987734205293489	0.987734205293489\\
5	24.8935582631618	1.01076117904739	1.01076117904739\\
6	22.4935218722059	1.1930521253715	1.1930521253715\\
7	22.2133639726158	1.19317580576616	1.19317580576616\\
8	17.2683805768305	0.966918238850981	0.966918238850981\\
9	16.0875717194243	0.981528165214495	0.981528165214495\\
10	15.5777799378473	0.90671532530801	0.90671532530801\\
11	15.2053681543714	0.855541211097943	0.855541211097943\\
12	14.964226772449	0.811978207415963	0.811978207415963\\
13	14.8014328026732	0.8816701033366	0.8816701033366\\
14	14.7790227035824	0.881790529299981	0.881790529299981\\
15	14.7793638174209	0.882577971091562	0.882577971091562\\
16	14.7484285700691	0.87271531854196	0.87271531854196\\
17	14.7102606314985	0.851487368911781	0.851487368911781\\
18	14.6944483948232	0.841094848120566	0.841094848120566\\
};

\addplot [color=red, line width=1.5pt]
 plot [error bars/.cd, y dir = both, y explicit]
 table[row sep=crcr, y error plus index=2, y error minus index=3]{%
1	47.0791136148533	0.538587132430922	0.538587132430922\\
2	39.4843028503756	1.41339019339346	1.41339019339346\\
3	27.7931771723963	0.852499346105613	0.852499346105613\\
4	26.5125962327223	0.902317386681203	0.902317386681203\\
5	25.4292509604901	0.967035057293879	0.967035057293879\\
6	23.8702982097065	1.08176102094209	1.08176102094209\\
7	23.1194233120739	1.11273020422102	1.11273020422102\\
8	22.5136314693961	1.07857315864165	1.07857315864165\\
9	21.3814047106927	0.875763854175257	0.875763854175257\\
10	20.6866687051021	0.796856141388719	0.796856141388719\\
11	19.5160115876345	0.680004587402164	0.680004587402164\\
12	17.9129960911833	0.64944695214663	0.64944695214663\\
13	17.1216032991092	0.789299515183014	0.789299515183014\\
14	17.053572243925	0.791826578874616	0.791826578874616\\
15	17.0111221613167	0.78559649214983	0.78559649214983\\
16	16.973051462093	0.777721166024387	0.777721166024387\\
17	16.8917041574017	0.755413924641834	0.755413924641834\\
18	16.8768839286114	0.747916048836231	0.747916048836231\\
};

\pgfplotsset{
	after end axis/.code={
		\node[black,above] at (axis cs:8.5, 10){Higher is better};
}}

\end{axis}
\end{tikzpicture}%

%% file: chapters/04_simulations/exp3/exp3_summary_pcp.tex
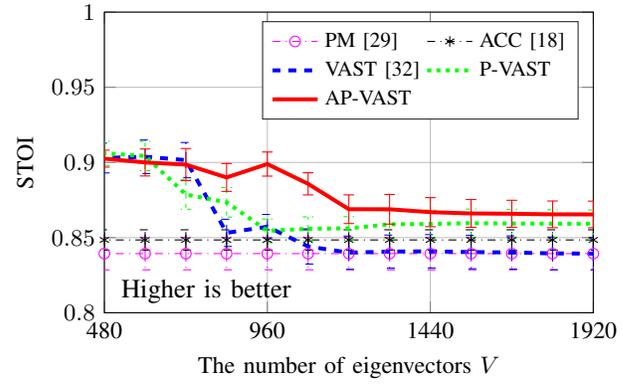
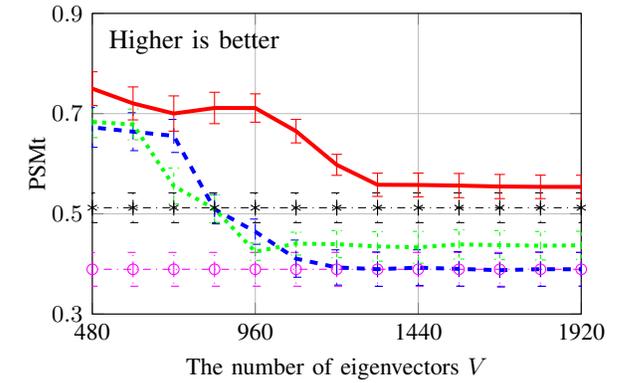
\begin{figure}[t]
	\centering
	\captionsetup[subfigure]{justification=centering}
		\subfloat[]{\input{chapters/04_simulations/exp3/exp3_all_free_stoi.tex}\label{subfig:exp3_stoi}}
		\quad
		\subfloat[]{\input{chapters/04_simulations/exp3/exp3_all_free_psmt.tex}\label{subfig:exp3_psmt}}
		
		\caption{\protect\subref{subfig:exp3_stoi} STOI in a range between $0.0$ and $1.0$ and \protect\subref{subfig:exp3_psmt} PSMt in a range between $-1.0$ to $1.0$ as a function of $V$ for five different methods: PM (purple dash-dot) \cite{Simon-Galvez2015}, ACC (black dash-dot) \cite{choi2002generation}, VAST (blue dash) \cite{Lee2018}, P-VAST (green dot), and AP-VAST (red solid). Note that both metrics are represented with the $95\%$ confidence interval (error bars).}
		\label{fig:exp3_results_pp}
\end{figure}

%% file: chapters/04_simulations/exp3/exp3_all_free_stoi.tex
%
%
\definecolor{mycolor2}{rgb}{0.00000,1.00000,1.00000}%
\definecolor{mycolor1}{rgb}{1.00000,0.00000,1.00000}%
\begin{tikzpicture}

\begin{axis}[%
width = 6.5 cm,
height = 4.0 cm,
scale only axis,
xmin=6,
xmax=18,
xtick={1,6,10,14,18},
xticklabels={{1},{480},{960},{1440},{1920}},
xlabel={The number of eigenvectors $V$},
ymin=0.8,
ymax=1.0,
ylabel={STOI},
axis background/.style={fill=white},
grid = major,
legend style={legend cell align=left, align=left, draw=white!15!black, font=\footnotesize},
legend columns = 2, legend pos = north east
]
\addplot [color=mycolor1, dashdotted, mark=o, mark options={solid, mycolor1}]
 plot [error bars/.cd, y dir = both, y explicit]
 table[row sep=crcr, y error plus index=2, y error minus index=3]{%
1	0.839360339749176	0.0107751830053662	0.0107751830053662\\
2	0.839360339749176	0.0107751830053662	0.0107751830053662\\
3	0.839360339749176	0.0107751830053662	0.0107751830053662\\
4	0.839360339749176	0.0107751830053662	0.0107751830053662\\
5	0.839360339749176	0.0107751830053662	0.0107751830053662\\
6	0.839360339749176	0.0107751830053662	0.0107751830053662\\
7	0.839360339749176	0.0107751830053662	0.0107751830053662\\
8	0.839360339749176	0.0107751830053662	0.0107751830053662\\
9	0.839360339749176	0.0107751830053662	0.0107751830053662\\
10	0.839360339749176	0.0107751830053662	0.0107751830053662\\
11	0.839360339749176	0.0107751830053662	0.0107751830053662\\
12	0.839360339749176	0.0107751830053662	0.0107751830053662\\
13	0.839360339749176	0.0107751830053662	0.0107751830053662\\
14	0.839360339749176	0.0107751830053662	0.0107751830053662\\
15	0.839360339749176	0.0107751830053662	0.0107751830053662\\
16	0.839360339749176	0.0107751830053662	0.0107751830053662\\
17	0.839360339749176	0.0107751830053662	0.0107751830053662\\
18	0.839360339749176	0.0107751830053662	0.0107751830053662\\
};
\addlegendentry{PM \cite{Simon-Galvez2015}}


\addplot [color=black, dashdotted, mark=asterisk, mark options={solid, black}]
 plot [error bars/.cd, y dir = both, y explicit]
 table[row sep=crcr, y error plus index=2, y error minus index=3]{%
1	0.848451055707347	0.00675456304411342	0.00675456304411342\\
2	0.848451055707347	0.00675456304411342	0.00675456304411342\\
3	0.848451055707347	0.00675456304411342	0.00675456304411342\\
4	0.848451055707347	0.00675456304411342	0.00675456304411342\\
5	0.848451055707347	0.00675456304411342	0.00675456304411342\\
6	0.848451055707347	0.00675456304411342	0.00675456304411342\\
7	0.848451055707347	0.00675456304411342	0.00675456304411342\\
8	0.848451055707347	0.00675456304411342	0.00675456304411342\\
9	0.848451055707347	0.00675456304411342	0.00675456304411342\\
10	0.848451055707347	0.00675456304411342	0.00675456304411342\\
11	0.848451055707347	0.00675456304411342	0.00675456304411342\\
12	0.848451055707347	0.00675456304411342	0.00675456304411342\\
13	0.848451055707347	0.00675456304411342	0.00675456304411342\\
14	0.848451055707347	0.00675456304411342	0.00675456304411342\\
15	0.848451055707347	0.00675456304411342	0.00675456304411342\\
16	0.848451055707347	0.00675456304411342	0.00675456304411342\\
17	0.848451055707347	0.00675456304411342	0.00675456304411342\\
18	0.848451055707347	0.00675456304411342	0.00675456304411342\\
};
\addlegendentry{ACC \cite{choi2002generation}}


\addplot [color=blue, dashed, line width=1.5pt]
 plot [error bars/.cd, y dir = both, y explicit]
 table[row sep=crcr, y error plus index=2, y error minus index=3]{%
1	0.758467773692744	0.0038385923501164	0.0038385923501164\\
2	0.855283531205436	0.00573863186854903	0.00573863186854903\\
3	0.888063204819416	0.00490700524103663	0.00490700524103663\\
4	0.918180981812047	0.0038367448435412	0.0038367448435412\\
5	0.926346595685974	0.00468653084879274	0.00468653084879274\\
6	0.902956384623614	0.00993137667835246	0.00993137667835246\\
7	0.903796866203914	0.0111538166124394	0.0111538166124394\\
8	0.901551619720193	0.0117240294128677	0.0117240294128677\\
9	0.853167118558318	0.00904156308013516	0.00904156308013516\\
10	0.857252052404613	0.00816967096118305	0.00816967096118305\\
11	0.844111913448271	0.0117012168372714	0.0117012168372714\\
12	0.839984661938139	0.0110591813238955	0.0110591813238955\\
13	0.840685923349381	0.0109966988882421	0.0109966988882421\\
14	0.840878993161187	0.0111102201015833	0.0111102201015833\\
15	0.840378904250101	0.0111026094627745	0.0111026094627745\\
16	0.840106187621744	0.0110527996073792	0.0110527996073792\\
17	0.83948745431361	0.0109119558034594	0.0109119558034594\\
18	0.839360339749176	0.0107751830053662	0.0107751830053662\\
};
\addlegendentry{VAST \cite{Lee2018}}

\addplot [color=green, dotted, line width=1.5pt]
 plot [error bars/.cd, y dir = both, y explicit]
 table[row sep=crcr, y error plus index=2, y error minus index=3]{%
1	0.797156334282089	0.00389222139156272	0.00389222139156272\\
2	0.873126589995677	0.00487773169159302	0.00487773169159302\\
3	0.893257989792644	0.00562140959909051	0.00562140959909051\\
4	0.899342625498905	0.00484180461757412	0.00484180461757412\\
5	0.923323008733697	0.00405392429531988	0.00405392429531988\\
6	0.906105153485755	0.00804907531965401	0.00804907531965401\\
7	0.904315483971867	0.00974818264920826	0.00974818264920826\\
8	0.878672464498723	0.00977552016471816	0.00977552016471816\\
9	0.873490643164599	0.00980989324059078	0.00980989324059078\\
10	0.854541594165206	0.00786189023842336	0.00786189023842336\\
11	0.855860792129322	0.00795086615215369	0.00795086615215369\\
12	0.856186830934503	0.00775559013188884	0.00775559013188884\\
13	0.858860902767748	0.0090403650988398	0.0090403650988398\\
14	0.859073198944587	0.00923239613425227	0.00923239613425227\\
15	0.859621143415319	0.00933782752371192	0.00933782752371192\\
16	0.859362294843786	0.00930733855779826	0.00930733855779826\\
17	0.859273180070791	0.00918557430062738	0.00918557430062738\\
18	0.859213278376485	0.00912691767977353	0.00912691767977353\\
};
\addlegendentry{P-VAST}

\addplot [color=red, line width=1.5pt]
 plot [error bars/.cd, y dir = both, y explicit]
 table[row sep=crcr, y error plus index=2, y error minus index=3]{%
1	0.377838211780999	0.00132141122966002	0.00132141122966002\\
2	0.724926122960478	0.00196963171781874	0.00196963171781874\\
3	0.829494559099311	0.004371579250388	0.004371579250388\\
4	0.872047410688049	0.00307123336575321	0.00307123336575321\\
5	0.892120694920595	0.00464115807470513	0.00464115807470513\\
6	0.902537157243761	0.0056497160332803	0.0056497160332803\\
7	0.899984524183061	0.00896741230345144	0.00896741230345144\\
8	0.898569808345837	0.0106959747027428	0.0106959747027428\\
9	0.890022426072223	0.0092467594148336	0.0092467594148336\\
10	0.898932913865746	0.00801443293992053	0.00801443293992053\\
11	0.885779158924514	0.00737576261169306	0.00737576261169306\\
12	0.868883175810267	0.00954595831167901	0.00954595831167901\\
13	0.868844549867564	0.00986643385653544	0.00986643385653544\\
14	0.86693433612473	0.00965859198109819	0.00965859198109819\\
15	0.866083301832936	0.00923928373382166	0.00923928373382166\\
16	0.865856604096727	0.00898017031591722	0.00898017031591722\\
17	0.865482625114187	0.00887538495797278	0.00887538495797278\\
18	0.865429091040952	0.0088433092051016	0.0088433092051016\\
};
\addlegendentry{AP-VAST}

\pgfplotsset{
	after end axis/.code={
		\node[black,above] at (axis cs:8.5, 0.8){Higher is better};
}}

\end{axis}
\end{tikzpicture}%

%% file: chapters/04_simulations/exp3/exp3_all_free_psmt.tex
%
%
\definecolor{mycolor2}{rgb}{0.00000,1.00000,1.00000}%
\definecolor{mycolor1}{rgb}{1.00000,0.00000,1.00000}%
\begin{tikzpicture}

\begin{axis}[%
width = 6.5 cm,
height = 4.0 cm,
scale only axis,
xmin=6,
xmax=18,
xtick={1,6,10,14,18},
xticklabels={{1},{480},{960},{1440},{1920}},
ytick={0.3, 0.5, 0.7, 0.9},
xlabel={The number of eigenvectors $V$},
ymin=0.3,
ymax=0.9,
ylabel style={yshift=-1ex},
ylabel={PSMt},
axis background/.style={fill=white},
grid = major,
legend style={legend cell align=left, align=left, draw=white!15!black, font=\footnotesize},
legend columns = 2, legend pos = north east
]
\addplot [color=mycolor1, dashdotted, mark=o, mark options={solid, mycolor1}]
 plot [error bars/.cd, y dir = both, y explicit]
 table[row sep=crcr, y error plus index=2, y error minus index=3]{%
1	0.389278933859359	0.0337439679896014	0.0337439679896014\\
2	0.389278933859359	0.0337439679896014	0.0337439679896014\\
3	0.389278933859359	0.0337439679896014	0.0337439679896014\\
4	0.389278933859359	0.0337439679896014	0.0337439679896014\\
5	0.389278933859359	0.0337439679896014	0.0337439679896014\\
6	0.389278933859359	0.0337439679896014	0.0337439679896014\\
7	0.389278933859359	0.0337439679896014	0.0337439679896014\\
8	0.389278933859359	0.0337439679896014	0.0337439679896014\\
9	0.389278933859359	0.0337439679896014	0.0337439679896014\\
10	0.389278933859359	0.0337439679896014	0.0337439679896014\\
11	0.389278933859359	0.0337439679896014	0.0337439679896014\\
12	0.389278933859359	0.0337439679896014	0.0337439679896014\\
13	0.389278933859359	0.0337439679896014	0.0337439679896014\\
14	0.389278933859359	0.0337439679896014	0.0337439679896014\\
15	0.389278933859359	0.0337439679896014	0.0337439679896014\\
16	0.389278933859359	0.0337439679896014	0.0337439679896014\\
17	0.389278933859359	0.0337439679896014	0.0337439679896014\\
18	0.389278933859359	0.0337439679896014	0.0337439679896014\\
};


\addplot [color=black, dashdotted, mark=asterisk, mark options={solid, black}]
 plot [error bars/.cd, y dir = both, y explicit]
 table[row sep=crcr, y error plus index=2, y error minus index=3]{%
1	0.512177825414226	0.0297601520608552	0.0297601520608552\\
2	0.512177825414226	0.0297601520608552	0.0297601520608552\\
3	0.512177825414226	0.0297601520608552	0.0297601520608552\\
4	0.512177825414226	0.0297601520608552	0.0297601520608552\\
5	0.512177825414226	0.0297601520608552	0.0297601520608552\\
6	0.512177825414226	0.0297601520608552	0.0297601520608552\\
7	0.512177825414226	0.0297601520608552	0.0297601520608552\\
8	0.512177825414226	0.0297601520608552	0.0297601520608552\\
9	0.512177825414226	0.0297601520608552	0.0297601520608552\\
10	0.512177825414226	0.0297601520608552	0.0297601520608552\\
11	0.512177825414226	0.0297601520608552	0.0297601520608552\\
12	0.512177825414226	0.0297601520608552	0.0297601520608552\\
13	0.512177825414226	0.0297601520608552	0.0297601520608552\\
14	0.512177825414226	0.0297601520608552	0.0297601520608552\\
15	0.512177825414226	0.0297601520608552	0.0297601520608552\\
16	0.512177825414226	0.0297601520608552	0.0297601520608552\\
17	0.512177825414226	0.0297601520608552	0.0297601520608552\\
18	0.512177825414226	0.0297601520608552	0.0297601520608552\\
};


\addplot [color=blue, dashed, line width=1.5pt]
 plot [error bars/.cd, y dir = both, y explicit]
 table[row sep=crcr, y error plus index=2, y error minus index=3]{%
1	0.594477861886006	0.00753940555670601	0.00753940555670601\\
2	0.830684592575505	0.00464531578338437	0.00464531578338437\\
3	0.791011630241185	0.0187976321977207	0.0187976321977207\\
4	0.739955275761012	0.0381182070681155	0.0381182070681155\\
5	0.737071262669234	0.0399310059333512	0.0399310059333512\\
6	0.672603066011615	0.0398377091318142	0.0398377091318142\\
7	0.663970855622781	0.0378808498754205	0.0378808498754205\\
8	0.6554752547299	0.0329844023615801	0.0329844023615801\\
9	0.50908104481007	0.0291633094854235	0.0291633094854235\\
10	0.464384747226595	0.0251770775869746	0.0251770775869746\\
11	0.410541201261728	0.0370954572077295	0.0370954572077295\\
12	0.392974854821447	0.0351699959293969	0.0351699959293969\\
13	0.389485443024088	0.0343528042123195	0.0343528042123195\\
14	0.392562580429388	0.0356302976267862	0.0356302976267862\\
15	0.390092894213027	0.0345288914214055	0.0345288914214055\\
16	0.38748405179836	0.033548014330807	0.033548014330807\\
17	0.389509526148723	0.0342264133179441	0.0342264133179441\\
18	0.389278933859359	0.0337439679896014	0.0337439679896014\\
};

\addplot [color=green, dotted, line width=1.5pt]
 plot [error bars/.cd, y dir = both, y explicit]
 table[row sep=crcr, y error plus index=2, y error minus index=3]{%
1	0.36269526068158	0.0201055719369408	0.0201055719369408\\
2	0.83410343772335	0.00396536880348342	0.00396536880348342\\
3	0.813094164732465	0.0134397898341838	0.0134397898341838\\
4	0.804083388644823	0.019564437469567	0.019564437469567\\
5	0.756298428501708	0.0248291537479187	0.0248291537479187\\
6	0.683433020103533	0.0313498469725611	0.0313498469725611\\
7	0.678358999243608	0.0306322783626119	0.0306322783626119\\
8	0.555443399991484	0.0359535471703669	0.0359535471703669\\
9	0.509843085369148	0.027680792059032	0.027680792059032\\
10	0.424432875894414	0.0172977469183871	0.0172977469183871\\
11	0.440563081941709	0.0227438752745545	0.0227438752745545\\
12	0.439468562104827	0.0270981623294893	0.0270981623294893\\
13	0.43486256653599	0.0293716312731985	0.0293716312731985\\
14	0.433200923912243	0.0317651290968909	0.0317651290968909\\
15	0.438893241585906	0.0291341751510091	0.0291341751510091\\
16	0.437545789789144	0.0290212068313211	0.0290212068313211\\
17	0.436290804443196	0.0283828326698639	0.0283828326698639\\
18	0.437122220335089	0.0281237237512034	0.0281237237512034\\
};

\addplot [color=red, line width=1.5pt]
 plot [error bars/.cd, y dir = both, y explicit]
 table[row sep=crcr, y error plus index=2, y error minus index=3]{%
1	0.124315752990408	0.00350587099602675	0.00350587099602675\\
2	0.805526223781089	0.00368958432510378	0.00368958432510378\\
3	0.7743426209482	0.0101796919798208	0.0101796919798208\\
4	0.773670051557112	0.0187949440374031	0.0187949440374031\\
5	0.776588414825642	0.0268125174974928	0.0268125174974928\\
6	0.749850919918626	0.0336490148519341	0.0336490148519341\\
7	0.720357080437978	0.0328830137067151	0.0328830137067151\\
8	0.700143548005496	0.0351846218325211	0.0351846218325211\\
9	0.711255387220316	0.0313605958673934	0.0313605958673934\\
10	0.711127008286766	0.0282868582925804	0.0282868582925804\\
11	0.664978326036003	0.0234033231389224	0.0234033231389224\\
12	0.597758215552322	0.0209609880514714	0.0209609880514714\\
13	0.557961102277973	0.0232108095105712	0.0232108095105712\\
14	0.557603416465537	0.0236716561472649	0.0236716561472649\\
15	0.556354434111545	0.024168201885433	0.024168201885433\\
16	0.554475713714365	0.0241282075235418	0.0241282075235418\\
17	0.553840370100754	0.0234068375907087	0.0234068375907087\\
18	0.553719222211201	0.0232900973965414	0.0232900973965414\\
};

\pgfplotsset{
	after end axis/.code={
		\node[black,above] at (axis cs:8.5, 0.8){Higher is better};
}}

\end{axis}
\end{tikzpicture}%

%% file: chapters/04_simulations/exp4/exp4_summary.tex
\begin{table*}[t]
\begin{threeparttable}
\caption{The Mean and the $95\%$ Confidence Interval of the Performance Metrics of the Speech Signal for Zone $\beta$ in the Reverberant Environment}
\label{tab:exp4_pfmmtx}

\begin{tabular*}{2\columnwidth}{@{\extracolsep{\fill}} r c c r r r r r}
\toprule
     Method & \multicolumn{2}{c}{Parameter} & \multicolumn{5}{c}{Performance metric}\\ 
\cmidrule{2-3} 
    & \multicolumn{1}{c}{$1\leq V \leq 1920$} & \multicolumn{1}{c}{$\mu \geq 0$} & \multicolumn{1}{c}{AC (dB)} & \multicolumn{1}{c}{nSDP (dB)} & {TIR (dB)} & \multicolumn{1}{c}{STOI} & \multicolumn{1}{c}{PSMt}\\
\midrule
    No control & {NA\tnote{a}} & NA & $-1.3$ & $14.2 \pm 0.5$  & $-0.1 \pm 0.3$ & $0.64 \pm 0.004$ & $0.14 \pm 0.025$\\
\addlinespace
    PM \cite{Simon-Galvez2015} & $1920$ & $1.0$ & $10.5$ & $\mathbf{-9.9 \pm 0.7}$ & $11.5 \pm 0.5$ & $0.80 \pm 0.003$ & $0.39 \pm 0.009$\\
    ACC \cite{choi2002generation} & NA & NA & $9.2$ & $0.4 \pm 0.8$ & $9.3 \pm 0.4$ & $0.77 \pm 0.013$ & $0.40 \pm 0.014$\\
\addlinespace
    VAST \cite{Lee2018} & $1080$ & $1.0$ & $12.2$ & $-7.8 \pm 0.7$ & $\mathbf{12.8} \pm \mathbf{0.7}$ & $0.80 \pm 0.004$ & $0.43 \pm 0.016$\\
    P-VAST & $1080$ & $1.0$ & $\mathbf{14.9}$ & $-7.2 \pm 0.6$ & $11.8 \pm 0.6$ & $0.76 \pm 0.004$ & $0.40 \pm 0.022$\\
\addlinespace
    AP-VAST & $1080$ & $1.0$ & $12.0$ & $-8.4 \pm 0.4$ & $12.2 \pm 0.3$ & $\mathbf{0.82} \pm \mathbf{0.004}$ & $\mathbf{0.54} \pm \mathbf{0.010}$\\
\bottomrule
\end{tabular*}

\scriptsize
\begin{tablenotes}
\RaggedRight
\item[a] NA: Not applicable
\end{tablenotes}
\end{threeparttable}
\end{table*}



%% file: chapters/04_simulations/ch4_1_comp_complexity.tex
\subsection{Computational complexity and processing time}
\label{ssec:discussion}
\input{chapters/04_simulations/exp0/exp0_comcomplx.tex}

The experiments showed that AP-VAST outperformed the reference methods in terms of the perceptual metrics. Thus, there is a clear benefit to making a sound zone control algorithm signal-adaptive and perceptually optimized. The price for this, however, is a higher computational complexity since we must update the control filters for every segment instead of just once for all segments. To quantify this, we here use the big-O notation $\mathcal{O}$ from \cite{Golub2013matrix} to denote the computational complexity of an algorithm. By using this notation, the joint diagonalization in AP-VAST, P-VAST, and VAST has a computational complexity of order $\mathcal{O}(L^3J^3)$ where $L$ and $J$ are the number of loudspeakers and the control filter length, respectively. Since the joint diagonalization has to be performed for every segment in AP-VAST, the resulting complexity is $\mathcal{O}(IL^3J^3)$ where $I$ is the number of segments. The joint diagonalization is performed from the spatial statistics. If the spatial statistics \rr{are} computed na\"ively, the computational complexity is $\mathcal{O}(MNL^2J^2)$ for every segment where $M$ and $N$ are the number of control points and the segment length, respectively. \mjr{Note that the complexity becomes high for P-VAST if $N$ becomes large.} The broadband PM and ACC-PM, on the other hand, demand the same order of complexity as the joint diagonalization, i.e., $\mathcal{O}(L^3J^3)$ to solve a large least-squares problem. Since the ACC method is solved in the frequency-domain, we get many smaller problems \rr{rather than one big problem}, one for every frequency bin, {which} results in a complexity of $\mathcal{O}(L^3)$ for solving the generalized eigenvalue problem and a complexity of $\mathcal{O}(L^2)$ for forming the spatial statistics in one of $J$ frequency bins. \rr{The above discussion} is summarized in Table~\ref{tab:exp0_cc}.

\input{chapters/04_simulations/processing_time/ptime_summary.tex}

\input{chapters/05_listeningtest/relevant/excerpts.tex}

\mjr{Lastly, the mean processing times for calculating the spatial correlation matrix and the joint diagonalization by AP-VAST are shown in Fig.~\ref{fig:ptime_summary}. Note that the $95\%$ confidence interval was negligible compared to the processing time. The same setup and the input signals used in Sec.~\ref{ssec:sim1_performEvaluation} except for the number of loudspeakers were used for computing the processing times. Four different numbers of loudspeakers were chosen, $L = \{4, 8, 12, 16\}$; therefore, the corresponding dimensions of the spatial correlation matrices are $LJ = \{960, 1920, 2880, 3840\}$, respectively, because the length of control filters is $J = 240$. All timings were computed on a Windows~10 desktop PC (Dell OptiPlex 5040) with a $3.4$~GHz Intel(R) Core(TM) i$7$-$6700$ CPU and $8$~GB RAM using the function in \textsc{Matlab}~2019a called \texttt{timeit}. As can be seen from Fig.~\ref{fig:ptime_summary}, when $LJ = 960$ for a $60$~ms time segment, the mean processing times are approximately $647$~ms and $88$~ms, respectively, and they are nearly $6027$~ms and $7484$~ms when $LJ = 3840$, respectively. The processing time for computing the joint diagonalization grows approximately eight times every time $LJ$ doubles due to the cubic complexity, as summarized in Table~\ref{tab:exp0_cc}. It should be noted that real-time in a practical setup would be challenging due to its substantial computational complexity.} 

%% file: chapters/04_simulations/exp0/exp0_comcomplx.tex
\begin{table}[t]
\begin{threeparttable}
\caption{The Computational Complexity for the Calculation of Spatial Statistics and Control Filters}
\label{tab:exp0_cc}

\begin{tabular*}{\columnwidth}{@{\extracolsep{\fill}} r l l}
\toprule
     Method & \multicolumn{1}{c}{Spatial statistics} & \multicolumn{1}{c}{Control filter}\\ 
\midrule
    PM \cite{Simon-Galvez2015} & $\mathcal{O}(ML^2J^2)$ & $\mathcal{O}(L^3J^3)$ for solving the least squares\\
    ACC \cite{choi2002generation} & $\mathcal{O}(L^2J)$ & $\mathcal{O}(L^3J)$ for solving $J$ GEPs\\
\addlinespace
    VAST \cite{Lee2018} & $\mathcal{O}(MNL^2J^2)$ & $\mathcal{O}(L^3J^3)$ for computing JD \\
    P-VAST & $\mathcal{O}(MNL^2J^2)$ & $\mathcal{O}(L^3J^3)$ for computing JD \\
\addlinespace
    AP-VAST & $\mathcal{O}(IMNL^2J^2)$ & $\mathcal{O}(IL^3J^3)$ for computing $I$ JDs \\
\bottomrule
\end{tabular*}
\smallskip
\begin{tablenotes}[leftmargin = 0.8in]
\RaggedRight
    \item GEP: generalized eigenvalue problem
    \item JD: joint diagonalization
\end{tablenotes}
\end{threeparttable}
\end{table}

%% file: chapters/04_simulations/processing_time/ptime_summary.tex
\begin{figure}[t]
	\centering
	\captionsetup[subfigure]{justification=centering}
		\subfloat[]{\input{chapters/04_simulations/processing_time/ptime_spacor.tex}\label{subfig:ptime_spacor}}\quad
		\subfloat[]{\input{chapters/04_simulations/processing_time/ptime_jdiag.tex}\label{subfig:ptime_jdiag}}
		
		\caption{The processing time of AP-VAST for calculating \protect\subref{subfig:ptime_spacor} The spatial correlation matrix $\vt{R}_\text{C}$ and \protect\subref{subfig:ptime_jdiag} The joint diagonalization. Four different numbers of loudspeakers $L = \{4,8,12,16\}$ with a fixed control filter length $J = 240$ lead us to have four different dimensions $LJ$ of $\vt{R}_\text{C}$: $LJ = \{960, 1920, 2880, 3840\}$.}
		\label{fig:ptime_summary}
\end{figure}
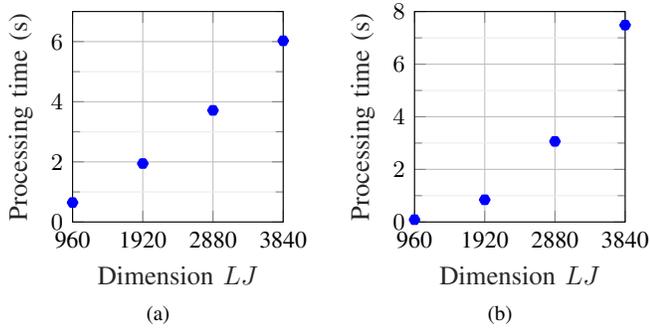

%% file: chapters/04_simulations/processing_time/ptime_spacor.tex
\begin{tikzpicture}

\begin{axis}[%
width=2.8cm,
height=2.8cm,
scale only axis,
xmin=1,
xmax=4,
xtick={1,2,3,4},
xticklabels={{960},{1920},{2880},{3840}},
xlabel style={font=\color{white!15!black}},
xlabel={Dimension $LJ$},
ymin=0,
ymax=7,
ylabel style={font=\color{white!15!black}},
ylabel={Processing time (s)},
grid = both,
xmajorgrids,
ymajorgrids,
minor y tick num = 1,
major grid style = {lightgray},
minor grid style = {lightgray!30},
]
\addplot [color=blue, line width=2.0pt, draw=none, mark=asterisk, mark options={solid, blue}]
 plot [error bars/.cd, y dir = both, y explicit]
 table[row sep=crcr, y error plus index=2, y error minus index=3]{%
1	0.6465119563	0.00204068078890712	0.00204068078890712\\
2	1.9443446113	0.0113947583529787	0.0113947583529787\\
3	3.7120753813	0.0131253502489667	0.0131253502489667\\
4	6.0267170513	0.0403722037028335	0.0403722037028335\\
};


\end{axis}
\end{tikzpicture}%

%% file: chapters/04_simulations/processing_time/ptime_jdiag.tex
\begin{tikzpicture}

\begin{axis}[%
width=2.8cm,
height=2.8cm,
scale only axis,
xmin=1,
xmax=4,
xtick={1,2,3,4},
xticklabels={{960},{1920},{2880},{3840}},
xlabel style={font=\color{white!15!black}},
xlabel={Dimension $LJ$},
ymin=0,
ymax=8,
ylabel style={font=\color{white!15!black}},
ylabel={Processing time (s)},
grid = both,
xmajorgrids,
ymajorgrids,
minor y tick num = 1,
major grid style = {lightgray},
minor grid style = {lightgray!30},
]

\addplot [color=blue, line width=2.0pt, draw=none, mark=asterisk, mark options={solid, blue}]
 plot [error bars/.cd, y dir = both, y explicit]
 table[row sep=crcr, y error plus index=2, y error minus index=3]{%
1	0.0877997363	0.00223002885411901	0.00223002885411901\\
2	0.8448418913	0.0187709300939541	0.0187709300939541\\
3	3.0632542863	0.0158897379549432	0.0158897379549432\\
4	7.4840307463	0.0322802101265303	0.0322802101265303\\
};

\end{axis}
\end{tikzpicture}%

%% file: chapters/05_listeningtest/relevant/excerpts.tex
\begin{table}[t]
\begin{threeparttable}
\caption{List of Excerpts Used in the MUSHRA tests}
\label{tab:excerpts}

\begin{tabular*}{\columnwidth}{@{\extracolsep{\fill}} c c c l} 
\toprule
     \multicolumn{1}{c}{\textbf{Scenario}} & \multicolumn{1}{c}{\textbf{Data set}} & \multicolumn{1}{c}{\textbf{Zone}} & \multicolumn{1}{c}{\textbf{Excerpt}}\\ 
\midrule
    \multirow{2}*{S1} & D1 & $\alpha$ & Female speech, Track 49\tnote{a}\\
    & D2 & $\beta$ & Male speech, Track 50\tnote{a}\\
\midrule
    \multirow{2}*{S2} & D3 & $\alpha$ & Pop music, Track 69\tnote{a}\\
    & D4 & $\beta$ & Pop music, Track 70\tnote{a}\\
\midrule
    \multirow{2}*{S3} & D5 & $\alpha$ & Piano solo\\
    & D6 & $\beta$ & Orchestra, Track 66\tnote{a}\\
\midrule
    \multirow{2}*{S4} & D7 & $\alpha$ & Guitar solo\\
    & D8 & $\beta$ & Male speech\\
\midrule
    \multirow{2}*{S5} & D9 & $\alpha$ & Zootopia dialogue in Danish\\
    & D10 & $\beta$ & Zootopia dialogue in English\\
\bottomrule
\end{tabular*}
\begin{tablenotes}[leftmargin = 0.8in]
\RaggedRight
    \item[a] EBU SQAM in \cite{ebusqam2008}
\end{tablenotes}
\end{threeparttable}
\end{table}

%% file: chapters/05_listeningtest/ch5_listening_test.tex
\mjr{
\subsection{Formal listening Test}
\label{ssec:listening_test}
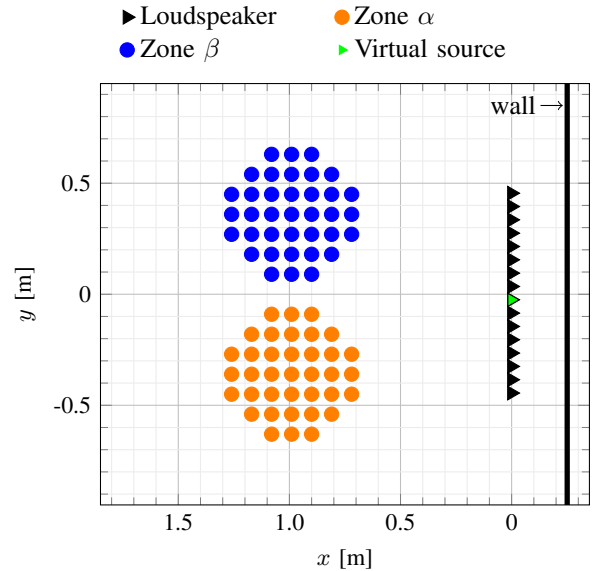
\begin{figure}[bt!]
	\centering
	\input{figures/systemGeometry_mushra.tex}
	\caption{The system geometry for the MUSHRA listening test. The number of loudspeakers (black triangle) $L=16$, the number of control points (orange and blue circles) $M=37$ for each zone, and the virtual source (green triangle) are shown.}
	\label{fig:experimentalsetup_mushra}
\end{figure}

In order to quantify the perceived performance of the candidate methods, we conducted a subjective listening test. The case of two bright zones was considered for five different scenarios. This led us to have ten different data sets\footnote{\mjr{The audio examples of the reproduced signals are available online at the following link: \url{https://tinyurl.com/apvast2020}}}, as summarized in Table~\ref{tab:excerpts}. In other words, the two reproduced sound fields were superposed; therefore, interference is present in the processed signal. In the listening test, the signal at the center of each zone was played back. Through this listening test, we evaluated the overall preference, i.e., the quality and the attenuation of the leakage to the other zone, of the candidate methods.

\subsubsection{Set-up}
\label{sssec:lt_setup}
As illustrated in Fig.~\ref{fig:experimentalsetup_mushra}, a uniform linear array of $16$ equally-spaced loudspeakers was considered. Each zone, which consists of $37$ control points, was located in front of the loudspeaker array. The distance between the loudspeakers was $9$~cm. The same distance was used for the space between the control points in each of the zones. A room with the dimensions $4.5\text{~m} \times 4.5\text{~m} \times 2.2\text{~m}$ and a reverberation time $T_{60}=300$~ms was considered. The measured RIRs used in \cite{Schneider2017, Schneider2019} were used for the listening tests. The impulse response of the desired sound field $h_{mz}[n]$ for the bright zone was chosen from the RIR of the $8$th loudspeaker, after being shortened to contain only the direct path component. The RIRs were resampled from $48$~kHz to $16$~kHz. Except for the above modifications, the same information described in Sec.~\ref{ssec:sim_systemsetup} was used. Therefore, $V$ is in a range of $1 \leq V \leq 3840 = LJ$.


The user parameters $V$ and $\mu$ were selected as $V = 3840$ and $\mu = 1$, respectively, for both AP-VAST and P-VAST. This choice allows us to directly compare the perceived performance of the perceptually optimized sound zones (AP-VAST and P-VAST) to that of the physically optimized sound zones (PM), as well as the perceived performance between AP-VAST and P-VAST.


\begin{figure}[bt!]
	\centering
	\input{chapters/05_listeningtest/relevant/mean_mos.tex}
	\caption{The mean values and the $95\%$ confidence intervals of all MUSHRA scores for four different methods and a hidden reference and the two anchors. In total, $1400$ ratings, specifically, $200$ ratings ($20$ participants for $10$ different data sets) per method, were used. Note that the standard anchor and the hidden reference are denoted as Anchor and Ref, respectively.}
	\label{fig:mean_mos_mushra}
\end{figure}
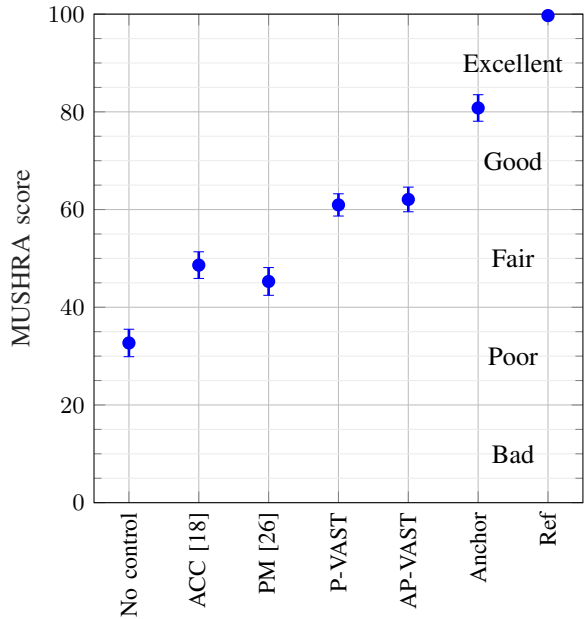

\subsubsection{MUSHRA test}
\label{sssec:lt_results}
A MUSHRA listening test was conducted according to the recommendation in \cite{itu15343mushra} using a webMUSHRA software \cite{Schoeffler2018}. In total, $20$~listeners with self-reported normal-hearing have participated in the tests. The listeners were asked to be located in a quiet place with wearing a pair of headphones\footnote{\mjr{We conducted the listening tests using an online platform \cite{Schoeffler2018} due to COVID-19 lockdown.}}. The listening test was divided into a training phase and an evaluation phase. In the training phase, the listeners were asked to get familiar with the interface and all the processed signals. In the evaluation phase, the listeners had to rate seven differently processed signals, the so-called stimuli, per data set in a range of $0$ to $100$ according to the quality with respect to the known reference. These stimuli included the hidden reference and a standard anchor (the low-pass filtered version of the reference at $3.5$~kHz). The hidden reference was used for examining the consistency of the listener's responses. The no control version was used as an additional anchor. The remaining methods were ACC, PM, P-VAST, and AP-VAST. In the test, they were displayed in random order. Therefore, one listener had to give $70$ ratings ($10$ data sets, $7$ stimuli) in total. 

\input{chapters/05_listeningtest/relevant/wcx_all.tex}

The statistical analyses at a significance level $\alpha$ of $0.05$ were conducted in SPSS 25 and \textsc{Matlab} 2019a. A Friedman test \cite{Friedman1937} was performed\footnote{\mjr{Initially, a one-way analysis of variance (ANOVA) with repeated measures was considered to compare the mean MUSHRA scores between ACC, PM, P-VAST, and AP-VAST because all the ratings by the same listener are on the same continuous, dependent variable, i.e., the MUSHRA scores.}}, as suggested in \cite{Mendonca2018}, because a Shapiro-Wilk test \cite{Shapiro1965, Royston1995} found that the MUSHRA scores were not normally distributed. It should be noted that the reference and the two anchors are excluded in the analyses. From the results of the Friedman test, a statistically significant difference was found in the perceived performance across all data sets between the related groups (ACC, PM, P-VAST, and AP-VAST), $N_\text{F} = 200, \chi^2(3)=183.799$, $p<0.001$ where $N_\text{F}$ is the number of ratings per method, $\chi^2(d)$ is the Friedman's Chi-square statistic with $d$ degree of freedom, and $p$ is the significance of the result. It can be interpreted as at least one pairwise difference is present, which is not surprising at all, considering Fig.~\ref{fig:mean_mos_mushra}. Therefore, post hoc analysis with Wilcoxon signed-rank tests \cite{Gibbons2011, Rey2011} was separately conducted. A Bonferroni correction was applied, resulting in a corrected $\alpha = 0.05/5 = 0.01$. We can observe the statistically significant pairwise difference ($p < 0.001$) between all the combinations, as summarized in Table~\ref{table:wilcoxon_sr_test_all}, except for the pair of P-VAST and AP-VAST.

We can visually see this difference by using a plot that shows the mean MUSHRA scores with the $95\%$ confidence intervals, as shown in Fig.~\ref{fig:mean_mos_mushra}. First, we can observe that the lowest mean MUSHRA score ($32.7$) when no control is considered. One of the significant observations is that the perceptual approaches (AP-VAST and P-VAST) outperform the existing methods (ACC and PM) by at least $10$ points in general, which is more than a $20\%$ improvement. Specifically, a significant improvement, which is more than $15$ points, is observed by comparing the scores between AP-VAST and PM. The mean MUSHRA scores for the four methods were as follows: $48.63$ for ACC, $45.29$ for PM, $60.95$ for P-VAST, and $62.07$ for AP-VAST. We emphasize that such a perceived difference was achieved even in the worst case of AC, when $V = LJ$ for a fixed $\mu$, for AP-VAST and P-VAST. Secondly, the standard anchor obtained about the mean MUSHRA score of $80.8$, which is the second-highest score. This observation can be interpreted that most listeners preferred the situation in which no interference is present, i.e., TIR is infinite, even though the processed signal is low-pass filtered. In other words, higher TIR and/or AC is preferred even the distortion (SDP) is present. This tendency seems more noticeable when speech is the desired signal than the case of music. Thirdly, the mean MUSHRA score by AP-VAST is slightly higher than that by P-VAST. 


\input{chapters/05_listeningtest/relevant/wcx_indi.tex}

We conducted Wilcoxon signed-rank tests for all the data sets separately to identify the statistically significant pairwise difference between AP-VAST and P-VAST. A two-tailed paired $t$-test \cite[Ch. 8.4]{Ross2004} could not be applied because the pairwise difference was not normally distributed, as in the previous analysis. The null hypothesis $H_0$ is as follows: the perceived performance between AP-VAST and P-VAST is the same. The null hypothesis is rejected if $p < 0.05$; otherwise, the null hypothesis cannot be rejected due to a lack of evidence to reject it at the significant level $\alpha = 0.05$. The test statistics of all the Wilcoxon tests are summarized in Table~\ref{table:wilcoxon_sr_test_indi}. The null hypothesis $H_0$ is rejected for the five data sets (D5, D6, D8, D9, and D10), which show a statistically significant difference in the perceived performance between AP-VAST and P-VAST. Specifically, the higher MUSHRA scores were obtained by AP-VAST over P-VAST from the four data sets: D6, D8, D9, and D10. We believe that segment-dependent $V$ and $\mu$ based on certain design criteria for constraints rather than fixed $V$ and $\mu$ could lead the optimal performance of AP-VAST, which will be dependent on the statistics of the input signals and different acoustic environments.}

%% file: figures/systemGeometry_mushra.tex
%
\begin{tikzpicture}

\begin{axis}[%
width=6.5cm,
scale only axis,
xmin=-2.1,
xmax=0.1,
xlabel={$x$ [m]},
xtick={-2.25, -1.75, -1.25, -0.75, -0.25, 0.25},
xticklabels={2.0, 1.5, 1.0, 0.5, 0, 0.5},
ymin=1.4,
ymax=3.1,
ytick={0.75, 1.25, 1.75, 2.25, 2.75, 3.25, 3.75},
yticklabels={-1.5, -1.0, -0.5, 0, 0.5, 1.0, 1.5},
ylabel={$y$ [m]},
axis background/.style={fill=white},
axis equal,
grid=both,
minor x tick num = 4,
minor y tick num = 4,
major grid style = {lightgray},
minor grid style = {lightgray!30},
legend style={at={(axis cs:-2.05,3.25)}, anchor=south west, legend cell align=left, align=left, draw=none, /tikz/every even column/.append style={column sep=2em}}, 
legend columns = 2,
]
\addplot[only marks, mark=triangle*, mark options={rotate=270}, mark size = 3pt, color=black, fill=black] table[row sep=crcr]{%
x	y\\
-0.25	1.805\\
-0.25	1.865\\
-0.25	1.925\\
-0.25	1.985\\
-0.25	2.045\\
-0.25	2.105\\
-0.25	2.165\\
-0.25	2.225\\
-0.25	2.285\\
-0.25	2.345\\
-0.25	2.405\\
-0.25	2.465\\
-0.25	2.525\\
-0.25	2.585\\
-0.25	2.645\\
-0.25	2.705\\
};
\addlegendentry{Loudspeaker}

\addplot[only marks, mark=*, mark options={}, mark size=2.7pt, color=orange, fill=orange] table[row sep=crcr]{%
x	y\\
-1.15	1.62\\
-1.24	1.62\\
-1.33	1.62\\
-1.06	1.71\\
-1.15	1.71\\
-1.24	1.71\\
-1.33	1.71\\
-1.42	1.71\\
-0.97	1.8\\
-1.06	1.8\\
-1.15	1.8\\
-1.24	1.8\\
-1.33	1.8\\
-1.42	1.8\\
-1.51	1.8\\
-0.97	1.89\\
-1.06	1.89\\
-1.15	1.89\\
-1.24	1.89\\
-1.33	1.89\\
-1.42	1.89\\
-1.51	1.89\\
-0.97	1.98\\
-1.06	1.98\\
-1.15	1.98\\
-1.24	1.98\\
-1.33	1.98\\
-1.42	1.98\\
-1.51	1.98\\
-1.06	2.07\\
-1.15	2.07\\
-1.24	2.07\\
-1.33	2.07\\
-1.42	2.07\\
-1.15	2.16\\
-1.24	2.16\\
-1.33	2.16\\
};
\addlegendentry{Zone $\alpha$}

\addplot[only marks, mark=*, mark options={}, mark size=2.7pt, color=blue, fill=blue] table[row sep=crcr]{%
x	y\\
-1.15	2.34\\
-1.24	2.34\\
-1.33	2.34\\
-1.06	2.43\\
-1.15	2.43\\
-1.24	2.43\\
-1.33	2.43\\
-1.42	2.43\\
-0.97	2.52\\
-1.06	2.52\\
-1.15	2.52\\
-1.24	2.52\\
-1.33	2.52\\
-1.42	2.52\\
-1.51	2.52\\
-0.97	2.61\\
-1.06	2.61\\
-1.15	2.61\\
-1.24	2.61\\
-1.33	2.61\\
-1.42	2.61\\
-1.51	2.61\\
-0.97	2.7\\
-1.06	2.7\\
-1.15	2.7\\
-1.24	2.7\\
-1.33	2.7\\
-1.42	2.7\\
-1.51	2.7\\
-1.06	2.79\\
-1.15	2.79\\
-1.24	2.79\\
-1.33	2.79\\
-1.42	2.79\\
-1.15	2.88\\
-1.24	2.88\\
-1.33	2.88\\
};
\addlegendentry{Zone $\beta$}

\addplot[only marks, mark=triangle*, mark options={rotate=270}, mark size=2pt, color=green, fill=green] table[row sep=crcr]{%
x	y\\
-0.25	2.225\\
};
\addlegendentry{Virtual source}

\addplot [color=black, line width=2.0pt, forget plot]
  table[row sep=crcr]{%
0	0\\
-4.5	0\\
-4.5	4.5\\
0	4.5\\
0	0\\
};
\draw [->] (axis cs:-0.12, 3.1) -- (axis cs:-0.02, 3.1) node[right, anchor=east, inner sep=10pt] {wall};

\end{axis}

\end{tikzpicture}%

%% file: chapters/05_listeningtest/relevant/mean_mos.tex
\begin{tikzpicture}

\begin{axis}[%
width=6.5cm,
height=6.5cm,
scale only axis,
xmin=0.5,
xmax=7.5,
xtick={1,2,3,4,5,6,7},
xticklabels={{No control},{ACC \cite{choi2002generation}},{PM \cite{Poletti2008}},{P-VAST},{AP-VAST},{Anchor},{Ref}},
xticklabel style={rotate=90},
ymin=0,
ymax=100,
ylabel style={font=\color{white!15!black}},
ylabel={MUSHRA score},
axis background/.style={fill=white},
minor y tick num = 3,
major grid style = {lightgray},
minor grid style = {lightgray!30},
xmajorgrids,ymajorgrids,
yminorgrids,
]
\addplot [color=blue, line width=1.0pt, draw=none, mark=*, mark options={solid, blue}]
 plot [error bars/.cd, y dir = both, y explicit, error bar style={line width = 1pt,solid},]
 table[row sep=crcr, y error plus index=2, y error minus index=3]{%
1	32.69	2.80787068228954	2.80787068228954\\
2	48.625	2.72981566224512	2.72981566224512\\
3	45.285	2.84231550396651	2.84231550396651\\
4	60.95	2.27784208958432	2.27784208958432\\
5	62.07	2.52407989886168	2.52407989886168\\
6	80.79	2.71535840411868	2.71535840411868\\
7	99.7	0.354526945256081	0.354526945256081\\
};

\pgfplotsset{
	after end axis/.code={
		\node[black] at (axis cs:6.5, 10){Bad};
		\node[black] at (axis cs:6.5, 30){Poor};
		\node[black] at (axis cs:6.5, 50){Fair};
		\node[black] at (axis cs:6.5, 70){Good};
		\node[black] at (axis cs:6.5, 90){Excellent};
	}
}

\end{axis}
\end{tikzpicture}%

%% file: chapters/05_listeningtest/relevant/wcx_all.tex
\begin{table}[t!]
\begin{threeparttable}
\caption{The Wilcoxon Signed-Rank Test Statistics for Different Pairs of the Sound Zone Control Methods}
\label{table:wilcoxon_sr_test_all}
    \begin{tabular*}{\columnwidth}{c @{\extracolsep{\fill}} |c|c|c|c|c}
    \toprule
        \multirow{2}*{Pair\ } & PM & PM & ACC & ACC & P-VAST\\
        & P-VAST & AP-VAST & P-VAST & AP-PVAST & AP-VAST\\
    \midrule
    $Z\ $ & $-9.536$ & $-10.066$ & $-9.018$ & $-9.580$ & $-1.145$\\
    $p\ $ & $<0.001$ & $<0.001$ & $<0.001$ & $<0.001$ & $0.252$\\
    \bottomrule
    \end{tabular*}
\begin{tablenotes}
\footnotesize
\item $Z$ denotes the $Z$ satistic.
\end{tablenotes}
\end{threeparttable}
\end{table}

%% file: chapters/05_listeningtest/relevant/wcx_indi.tex
\begin{table}[t!]
\begin{threeparttable}
\caption{The Wilcoxon Signed-Rank Test Statistics Between AP-VAST and P-VAST for Different Data Sets}
\label{table:wilcoxon_sr_test_indi}
    \begin{tabular*}{\columnwidth}{r@{\extracolsep{\fill}}|c|c|c|c|c} 
    \toprule
        \textbf{Scenario\ } & S1 & S2 & S3 & S4 & S5\\
    \midrule
        \textbf{Dataset\ } & D1 & D3 & D5 & D7 & D9\\
    \addlinespace
        $Z\ $ & $-1.269$ & $-1.188$ & $-3.511$ & $-1.572$ & $-2.073$\\
        $p\ $ & $0.205$ & $0.235$ & \underline{$<0.001$} & $0.116$ & $\mathbf{0.038}$\\
    \midrule
        \textbf{Dataset\ } & D2 & D4 & D6 & D8 & D10\\
    \addlinespace
        $Z\ $ & $-0.763$ & $-0.982$ & $-3.287$ & $-2.833$ & $-2.199$\\
        $p\ $ & $0.445$ & $0.326$ & $\mathbf{0.001}$ & $\mathbf{0.005}$ & $\mathbf{0.028}$\\
    \bottomrule
    \end{tabular*}
\begin{tablenotes}
\footnotesize
\item (\textbf{boldface}): $H_0$ is rejected, and AP-VAST has a higher mean MUSHRA score than P-VAST. (\underline{underlined}): $H_0$ is rejected, and P-VAST has a higher mean MUSHRA score than AP-VAST. Otherwise, $H_0$ cannot be rejected due to insufficient evidence to reject it.
\end{tablenotes}
\end{threeparttable}
\end{table}

%% file: chapters/06_conclusion/ch6_conclusion.tex
\section{Conclusion}
\label{sec:conclusion}
In this paper, we proposed a signal-adaptive method for creating perceptually optimized sound zones by using variable span trade-off filters in the time-domain. This method was achieved by taking the characteristics of input signals and the human auditory system into account segment-wise. The characteristics of input signals were integrated into the spatial correlation matrices, and the human auditory system was quantified mathematically by using a psychoacoustic model. Masking thresholds were calculated by using this model from the given input signal and used as perceptual weighting filters applied to the input signals. To this end, it allowed us to shape the reproduction error perceptually so that the interference becomes less or ideally inaudible to the listener in a given zone according to the human auditory system. Exploiting the joint diagonalization of the spatial correlation matrices allowed us to have a flexible control filter that trades-off the acoustic contrast and the signal distortion. Through validations in both anechoic and reverberant environments, the performance in terms of the physical metrics -- AC, TIR, and nSDP -- as well as the perceptual metrics -- STOI and PSMt -- was measured. The performance across all metrics, zones, and input signals was reasonably consistent, all indicating that the proposed method provides a perceptually better reproduction of the desired sound field, even though the physical metrics are not necessarily better. \mjr{Lastly, through a MUSHRA listening test, it was verified that the perceptually optimized sound zones provide more than $20\%$ better perceived performance in terms of the mean MUSHRA score compared to the existing sound zone control methods.}

%% file: chapters/07_appendix/appendixA.tex
\subsection{Acoustic contrast}
Since the acoustic contrast $\gamma(\vt{q})$ is the ratio between the power in the bright and dark zones, it can be written as
\begin{align}
	\gamma(\vt{q}) &= \frac{M_\text{D}}{M_\text{B}}\frac{\vt{q}^T\vt{R}_\text{B}\vt{q}}{\vt{q}^T\vt{R}_\text{D}\vt{q}},
\end{align}
and if we plug $\vt{q}_\textup{P-VAST}(V,\mu)$ from \eqref{eq:pvast_confilter} into $\gamma(\vt{q})$, then it yields
\begin{align}
    \gamma\left(\vt{q}_\textup{P-VAST}(V,\mu)\right) &= \frac{M_\text{D}}{M_\text{B}}\frac{\vt{a}_\textup{P-VAST}^T(V,\mu)\vt{\Lambda}_V\vt{a}_\textup{P-VAST}(V,\mu)}{\vt{a}_\textup{P-VAST}^T(V,\mu)\vt{a}_\textup{P-VAST}(V,\mu)}\ .
    \label{eq:acousticcontrast}
\end{align}
If we consider $V$ and $V+1$, respectively, then we obtain
\begin{align}
    \gamma\left(\vt{q}_\textup{P-VAST}(V,\mu)\right) &= \frac{M_\text{D}}{M_\text{B}} \frac{\sum_{v=1}^{V}{|a_v|^2\lambda_v}}{\sum_{v=1}^{V}{|a_v|^2}},\label{eq:acv}\\
    \gamma\left(\vt{q}_\textup{P-VAST}(V+1,\mu)\right) &= \frac{M_\text{D}}{M_\text{B}} \frac{\sum_{v=1}^{V+1}{|a_v|^2\lambda_v}}{\sum_{v=1}^{V+1}{|a_v|^2}},\label{eq:acv1}
\end{align}
where $a_v$ \rr{is} the $v$th element in $\vt{a}_\textup{P-VAST}(V,\mu)$. Subtracting \eqref{eq:acv1} from \eqref{eq:acv} and reducing to common denominator lead us to have
\begin{align}
\gamma&\left(\vt{q}_\textup{P-VAST}(V,\mu)\right) - \gamma\left(\vt{q}_\textup{P-VAST}(V+1,\mu)\right) \nonumber\\
    &= \frac{M_\text{D}}{M_\text{B}} \frac{|a_{V+1}|^2\Big(\sum_{v=1}^{V}{|a_v|^2\lambda_v} - \lambda_{V+1}\sum_{v=1}^{V}{|a_v|^2}\Big)}{\sum_{v=1}^{V}{|a_v|^2}\sum_{v=1}^{V+1}{|a_v|^2}} \nonumber\\
    &= \frac{M_\text{D}}{M_\text{B}} \frac{|a_{V+1}|^2\Big\{\sum_{v=1}^{V}{|a_v|^2\big(\lambda_v-\lambda_{V+1}\big)}\Big\}}{\sum_{v=1}^{V}{|a_v|^2}\sum_{v=1}^{V+1}{|a_v|^2}} \label{eq:vvsubstract}.
\end{align}
Since $|a_v|^2$ and $|a_{V+1}|^2$ are nonnegative and $\lambda_v \geq \lambda_{V+1}$ (the equality holds when $\lambda_v = 0$), the acoustic contrast monotonically decreases for \rr{increasing} $V$, i.e., $\gamma\left(\vt{q}_\textup{P-VAST}(V,\mu)\right) \geq \gamma\left(\vt{q}_\textup{P-VAST}(V+1,\mu)\right)$.

\subsection{Signal distortion}
If we plug $\vt{q}_\textup{P-VAST}(V,\mu)$ from \eqref{eq:pvast_confilter} into \eqref{eq:errbr} and \eqref{eq:errdk}, we obtain
\begin{align}
    \tilde{\mathcal{S}}_\text{B}(\vt{q}_\textup{P-VAST}(V,\mu)) &= \tilde{\sigma}_d^2 - 2\tilde{\vt{r}}_\text{B}^T\vt{U}_V \vt{G}^{-1} \vt{U}_V^T\tilde{\vt{r}}_\text{B} \nonumber\\
    &\quad +\tilde{\vt{r}}_\text{B}^T\vt{U}_V \vt{G}^{-1} \vt{\Lambda}_V \vt{G}^{-1}\vt{U}_V^T\tilde{\vt{r}}_\text{B} \nonumber\\
        &= \tilde{\sigma}_d^2 - \sum_{v=1}^{V}{\frac{\lambda_v + 2\mu}{(\lambda_v+\mu)^2}|\vt{u}_v^T\tilde{\vt{r}}_\text{B}|^2},\label{eq:sbvmu}\\
    \tilde{\mathcal{S}}_\text{D}(\vt{q}_\textup{P-VAST}(V,\mu)) &= \tilde{\vt{r}}_\text{B}^T\vt{U}_V \vt{G}^{-2}\vt{U}_V^T\tilde{\vt{r}}_\text{B} \nonumber\\
    &= 
        \sum_{v=1}^{V}{\frac{|\vt{u}_v^T\tilde{\vt{r}}_\text{B}|^2}{(\lambda_v+\mu)^2}}\ ,
    \label{eq:sdvmu}
\end{align}
where $\vt{G} = \vt{\Lambda}_V+\mu\vt{I}_V$. Interestingly, we can observe that $\tilde{\mathcal{S}}_\text{B}(\vt{q}_\textup{P-VAST}(V,\mu))$ decreases and $\tilde{\mathcal{S}}_\text{D}(\vt{q}_\textup{P-VAST}(V,\mu))$ increases monotonically for \rr{increasing} $V$, respectively, because all variables in \eqref{eq:sbvmu} and \eqref{eq:sdvmu} are nonnegative. Finally, we plug \eqref{eq:sbvmu} and \eqref{eq:sdvmu} into \eqref{eq:costfcn}, then we obtain
\begin{align}
    \mathcal{L}(\vt{q}_\textup{P-VAST}(V,\mu)) &= \tilde{\sigma}_d^2 - \sum_{v=1}^{V}{\frac{|\vt{u}_v^T\tilde{\vt{r}}_\text{B}|^2}{\lambda_v+\mu}},
\end{align}
which decreases for \rr{increasing} $V$. Therefore, we can obtain the minimum reproduction error when all eigenvectors are used and $\mu$ is a positive value\rr{,} which means that the residual power in the dark zone is still being controlled. Note that we obtain the minimum signal distortion if $\mu = 0$.